\documentstyle[12pt]{article}

\makeatletter
\@addtoreset{equation}{section}
\makeatother

\def\theequation{\thesection.\arabic{equation}}

\newcommand{\be}{\begin{equation}}
\newcommand{\ee}{\end{equation}}
\newcommand{\eee}{\end{eqnarray}}
\newcommand{\bee}{\begin{eqnarray}}

\newcommand{\ga}{\alpha}
\newcommand{\gb}{\beta}
\newcommand{\gga}{\gamma}
\newcommand{\gd}{\delta}
\newcommand{\gl}{\lambda}

\newcommand{\gep}{\epsilon}
\newcommand{\gvep}{\varepsilon}
\newcommand{\gs}{\sigma}
\newcommand{\go}{\omega}

\newcommand{\tp}{\tilde{p}}
\newcommand{\tq}{\tilde{q}}
\newcommand{\ty}{\tilde{y}}

\begin{document}
\thispagestyle{empty}

\begin{flushright}
\vspace{1mm}
FIAN/TD/16--98\\
{June 1998}\\
\end{flushright}

\vspace{1cm}

\begin{center}
{\large\bf
HIGHER-SPIN GAUGE INTERACTIONS FOR MASSIVE MATTER FIELDS IN 3D AdS
SPACE-TIME}\\
\vglue 1  true cm
\vspace{2cm}
 S.~F.~Prokushkin
\footnote{e-mail: prok@td.lpi.ac.ru}
  and M.~A.~Vasiliev
\footnote{e-mail: vasiliev@td.lpi.ac.ru}  \\
\vspace{1cm}

I.E.Tamm Department of Theoretical Physics, Lebedev Physical Institute,\\
Leninsky prospect 53, 117924, Moscow, Russia
\vspace{1.5cm}
\end{center}

\begin{abstract}
\baselineskip .5 true cm
\noindent
A remarkable feature of the models with interactions exhibiting
 higher-spin (HS) gauge
symmetries in $d>2$ is that their most symmetric vacua require
(anti)-de Sitter
(AdS) geometry  rather than the flat one. In  striking parallelism
to what might be expected of $M$ theory HS gauge theories describe
infinite towers of fields of all spins and possess naturally
space-time SUSY and Chan-Paton type inner symmetries. In this paper,
we analyze at the level of the equations of
motion the simplest non-trivial HS model which describes HS gauge
interactions (on the top of the usual supergravitational and
(Chern-Simons) Yang-Mills
interactions) of massive spin-0 and spin-1/2 matter fields in $d=2+1$
AdS space-time. The parameter of mass of the matter fields is identified
with the vev of a certain auxiliary field in the model. The matter fields
are shown to be arranged into $d3$ $N=2$ massive hypermultiplets in certain
representations of $U(n)\times U(m)$ Yang-Mills gauge groups. Discrete
symmetries of the full system are studied, and the related $N=1$
supersymmetric truncations with $O(n)$ and $Sp(n)$ Yang-Mills symmetries
are constructed.
The simplicity of the model allows us to
elucidate some general properties of the HS models.
In particular, a new result, which can have interesting implications to
the higher-dimensional models, is that our model is shown to admit an
``integrating" flow that proves existence of a non-local
B\"acklund-Nicolai--type mapping to the free system.
\end{abstract}

\newpage

\section{Introduction}

The idea that higher-spin (HS) extensions of the ordinary
space-time (super)symmetries
can drive us to a fundamental unified theory is old enough
(see e.g. \cite{hist1, hist2, hist3} and references therein).
The main problem was \cite{AD1, AD2, AD3} how
to construct theories with the interactions
consistent with HS gauge symmetries. Although some encouraging
results have been obtained \cite{BBB1, BBB2, BBB3, BBB4, BBB5}
the problem remained with
the gravitational interaction of HS gauge fields \cite{AD1, AD2, AD3}
as well as with the powerful no-go statements \cite{nogo1, nogo2} that
led to a dominating opinion that consistent HS
gauge theories cannot be constructed. A way out was found in \cite{FV1},
where gauge invariant interactions of $d=4$ massless HS gauge fields
compatible with gravity have been constructed at the action level
in the cubic order in interactions, and in \cite{more},
where these results were extended to
all orders in interactions at the level of equations of motion
(see also \cite{rev} for a recent review and further references).

The most important reason which made HS gauge theories
invisible within any expansion near the
flat space \cite{AD1, AD2, AD3} is that HS gauge interactions
with unbroken HS gauge symmetries require \cite{FV1}
the cosmological constant to be non-zero because some of
the gauge-invariant interactions contain its negative powers.
Simultaneously, this property allows one to avoid
the no-go statements \cite{nogo1, nogo2} which claim that HS symmetries
cannot be seen in any $S$-matrix constructed in the flat space.
It is important to stress that this feature of the HS gauge theories
does not mean that the flat background is impossible but rather it means
that in the flat background the HS symmetries should necessarily be
broken (spontaneously). Together with another important fact that
HS gauge theories contain towers of massless fields with
infinitely increasing spins (which form multiplets of the HS symmetries)
this implies that any spontaneously broken phase
of a HS gauge theory should contain infinitely many
massive HS excitations similarly to string theories.

Nowadays a unified theory is identified with $M$ theory, a
hypothetical theory which reduces to $d=11$ supergravity in the
low-energy limit and gives rise
to superstring theories by virtue of compactification of extra
dimensions. Since superstring theories describe infinite
towers of massive HS excitations on the top of a finite
number of massless modes it is natural to expect
that $M$ theory might be some HS gauge theory
in eleven or higher dimensions and that ordinary string theories in
ten and lower dimensions result from the originally massless HS gauge
fields by means of spontaneous breakdown of
the HS gauge symmetries via compactification
of extra dimensions. Massless modes of the superstring theory
(supergravity) are then expected to
correspond to unbroken maximal
finite dimensional subalgebras of the original
infinite-dimensional HS gauge symmetry algebras. It is important to note that
HS gauge theories are fixed almost unambiguously by the gauge symmetry
principle.

{}From the most elaborated $d4$ example
it follows \cite{more,rev} that the ambiguity is
basically in the choice of the Yang-Mills (spin -1)
gauge group and is parallel to the ambiguity
in the Chan-Paton structure. The HS gauge theories contain only two
fundamental coupling constants, the gravitational constant $\kappa$
and the cosmological constant $\lambda$.
Their dimensionless combination $g^2 \sim \kappa^2\lambda^2$ is identified
with the Yang-Mills coupling constant (in $d=4$).
{}From the $d=3$ analysis of this paper
we will learn that some parameters of matter fields (masses) arise
as moduli of certain auxiliary scalar fields.

Recently, a remarkable conjecture was made
\cite{AdS1, FF, AdS2, AdS3, AdS4}
on the large $N$ correspondence of a fundamental theory
in AdS spaces to conformal theories on their boundaries.
This conjecture is an important development of the physics
of singletons \cite{singl1, singl2}. The degrees of freedom
of the singleton boundary models are identified with the
maximally supersymmetric conformal Yang-Mills models. A nature
of a fundamental bulk theory is more obscure. We believe that this
is a HS gauge theory in the AdS space.
This idea is somewhat reminiscent of the results of $d2$ analysis in
\cite{vafa}. A similar conjecture on the
HS gauge nature of $M$ theory was  recently put forward in \cite{sez}.

The HS symmetry algebras \cite{W}
were identified in \cite{OP1} with the infinite-dimensional Lie
superalgebras constructed from the
Heisenberg-Weyl algebras with spinorial generating elements
and some their further extensions \cite{KV1,OP2}
(see sections \ref{OpReal} and \ref{YM} for more details).
In other words, the HS symmetries are realized as (appropriately
supersymmetrized) Moyal star products \cite{MOY1, MOY2} in certain
auxiliary spinor (twistor) spaces. It is interesting that
the Moyal product was recently found to be relevant to the $M$ theory
in its M(atrix) formulation \cite{MM1, MM2, MM3, MM4}.
A parallelism between HS gauge
theories and the Fedosov quantization was recently emphasized
in \cite{Castro}. Another interesting parallelism
is due to the analysis of $N=2$ critical open superstring in \cite{DL}
where it was shown that physical degrees of freedom of the model
describe self-dual fields of arbitrary high spin and can naturally
be described in terms of a hyperspace with spinor commuting coordinates.
Remarkably, these additional coordinates have the same structure as
the auxiliary spinor variables used to describe the HS dynamics
in terms of the Moyal star product \cite{OP1}.

In this paper, we study the non-linear equations of motion
which describe HS interactions of massive matter fields in 2+1 dimensions.
Since HS gauge fields do not propagate in $d3$
\cite{AchTow, Witt, Blen}, the analysis of the $d3$ model
is simpler than that of the higher-dimensional models with propagating
HS fields. The simplicity of the model
allows us to make some of the  statements on its
properties rather explicit.
On the other hand, a general character of our conclusions
will allow us to speculate how these extend
to less trivial  models in higher dimensions.

We will demonstrate that the model at hand
exhibits a number of natural
properties in the context of possible applications to $M$ (string) theory.
In particular, we show that the full system possesses naturally
$N=2$ (target space) SUSY and describes HS interactions of $d3$
hypermultiplets. An essential property that fits nicely the
superstring picture is that the HS models constructed
in this paper admits $U(n)$ Yang-Mills (spin 1) gauge symmetries
 as well as $N=1$
supersymmetric truncations with orthogonal and symplectic gauge groups
(the sector of even spins, which contains gravity and
corresponds to closed string, is a
part of all HS gauge theories).
An important feature of the HS gauge theories  studied further
in this paper is that consistent HS gauge interactions require non-zero
cosmological constant. One of the main conclusions
is that the non-local character of the Moyal product together with the
presence of a non-zero cosmological constant might imply some
sort of non-locality of the HS models at the interaction level
because these
two properties allow HS interactions with arbitrary high derivatives
(but no nonlocalities at the linear level!). This is again in agreement
with what one would expect of a model underlying string theory
(in this respect the cosmological constant in HS gauge theories plays
a role analogous to
that of the string tension in superstring theory)
and has a striking similarity with the conclusions of the analysis of
the noncommutative Matrix models \cite{MM1, MM2, MM3, MM4}.

The main result of this paper consists of the explicit construction of
non-linear equations of motion which describe interactions of spin-0 and
spin-${1\over 2}$ massive matter fields via HS gauge potentials.
This model is a generalization of that with massless matter fields
proposed in \cite{Eq}. We show that the parameter of mass of the matter
fields appears as the vacuum value of a certain scalar auxiliary field
and construct a vacuum solution of the system which is
invariant under its discrete symmetries.
Another important result consists of the
explicit construction of a flow that
commutes with the original system and allows one to reduce constructively
its solutions to those of the linearized system by virtue
of a sort of B\"acklund-Nicolai non-linear map \cite{Nic}.
With the aid of the example of $d3$ gravity we demonstrate
that this mapping should be essentially non-local because
the relevant power series in higher derivatives of the original fields
are infinite beyond the linearized approximation.

The paper is self-contained and is
organized as follows. We start in sect.~\ref{Prel} with
the explanation of the formulation of the free field
equations in a form of covariant constancy conditions
supplemented with appropriate constraints. This formulation follows
\cite{Unf} and is adapted to theories with symmetries mixing
higher derivatives of the dynamical fields as it is the case in
the HS theories.
In sect.~\ref{OpReal} we reproduce the main results of \cite{BPV}
on the operator realization of free massive field equations.
In sect.~\ref{StarPr} an appropriate form of the star
(Moyal) product is defined and some of its properties are discussed.
In sect.~\ref{NonlSys} we present the full nonlinear
system which describes massive matter fields interacting via HS
gauge fields. Some vacuum solutions of this system are presented in
sect.~\ref{Vacuum}. In sect.~\ref{LinEq}
we analyze the linearization of the full system near the vacuum
solutions of sect.~\ref{Vacuum} and show that it reproduces
correctly the free field dynamics. In sect.~\ref{Integr}
we construct the integrating flow for the full nonlinear system
and discuss locality. In sect.~\ref{YM} we extend the full system
to the case with non-Abelian internal symmetries and study
reality conditions and truncations. In sect.~\ref{GlobSym}
we study global (super)symmetries of the system.
In Appendix A the regularity theorem is proved which guarantees
that all manipulations with the star product in this paper are
well defined. The details of the general construction of
the vacuum solution are considered in Appendix B.
In Appendix C we demonstrate the non-local character of
the ``integrating" transformations of sect.~\ref{Integr},
using the example of Einstein gravity.

\section{Preliminaries}\label{Prel}

It is well known \cite{geom1, geom2, geom3, geom4, geom5}
that geometry of space-time can be described in terms
of the connection 1-forms of an
appropriate space-time symmetry algebra $l$.
For example vielbein $h_{\nu}{}^a$ and Lorentz connection
$\go_{\nu}{}^{ab} = -\go_{\nu}{}^{ba}$
of the $d$-dimensional space-time
\footnote{ $\nu, \mu = 0,\ldots,(d-1)$ are indices of
1-forms while
$a,b \ldots = 0,\ldots,(d-1)$ are tangent indices
that are raised and lowered by the flat
Minkowski metrics $\eta^{ab} = diag(+- \cdots --)$.}
can be identified with the gauge fields of the Poincar\'e algebra with
the generators $P_a$ (translations) and $L_{ab}$ (Lorentz rotations),
\be
\label{Poinc}
   A = h^a P_a + \go^{ab} L_{ab} \,.
\ee
The Poincar\'e curvatures have a form
\be
\label{rim}
   R_{\mu\nu}{}^{ab}=\partial_{\mu}\go_{\nu}{}^{ab}
     -\partial_{\nu}\go_{\mu}{}^{ab}+\go_{\mu}{}^a{}_c\,
      \go_{\nu}{}^{cb}-\go_{\nu}{}^a{}_c\,\go_{\mu}{}^{cb}  \,,
\ee
\be
\label{tor}
   R_{\mu\nu}{}^a=\partial_{\mu}h_{\nu}{}^a-\partial_{\nu}h_{\mu}{}^a
     +\go_{\mu}{}^a{}_c h_{\nu}{}^c-\go_{\nu}{}^a{}_c h_{\mu}{}^c \,.
\ee
Assuming that $h_{\nu}{}^a$ is non-degenerate one identifies
$R_{\mu\nu}{}^{a}$ and $R_{\mu\nu}{}^{ab}$ with the torsion tensor and
the Riemann tensor
${\cal R}_{\mu\nu}{}^{ab}$,
respectively\footnote{We use conventions with
$R_{\mu\nu}{}^{ab} (\omega (h))$=
${\cal R}_{\mu\nu}{}^{ab}$ =
${\cal R}^{ab}{}_{\mu\nu}$=
$h_{\gl}{}^a
h^{\gs b} {\cal R}^\gl {}_{\gs\mu\nu}$, ${\cal R}^\gl {}_{\gs\mu\nu} =
\partial_{\mu}\Gamma^\gl_{\gs\nu}+ \Gamma^\rho_{\gs\nu}\Gamma^\gl_{\rho\mu}
- (\mu\leftrightarrow \nu)$. $\Gamma^\rho_{\gs\nu}$ is
 symmetric Christoffel connection defined via
$D_\nu g_{\mu\rho}=0$, and $\omega (h)$ solves $R_{\mu\nu}{}^a =0$.}.
The flat geometry is
described by the vielbein $h_{\nu}{}^a$ and the Lorentz connection
$\go_{\nu}{}^{ab}$ obeying the zero-curvature and zero-torsion conditions
\be
\label{zc}
  R_{\mu\nu}{}^{ab}=0\,,\quad R_{\mu\nu}{}^{a}=0 \,.
\ee
One can choose a solution of (\ref{zc}) in the form
\be
\label{gk}
h_{\nu}{}^a=\gd_{\nu}^a \,,\quad  \go_{\nu}{}^{ab}=0 \,.
\ee

Analogously, one can use gauge fields of the AdS
algebra $o(d-1,2)$ to describe the geometry of the AdS space-time.
Let us consider the $o(d-1,2)$ gauge fields $A_\mu^{BC}=-A_\mu^{CB}$
(the indices $B,C=0,...,d$ are raised and lowered by the flat
metrics $\eta^{BC} = diag(+- \cdots -+)$)
 and set $\go_{\mu}{}^{ab}=A_\mu^{ab}$,
$h_{\mu}{}^a =(\sqrt{2} \gl)^{-1} A_\mu^{a\, \cdot }$
with the conventions $a,b = 0,...,d-1$, $B=(b,\,\cdot)$.
Here $\gl \neq 0$ is some constant.
The respective $o(d-1,2)$ gauge curvatures have the form
\be
\label{rim1}
   R_{\mu\nu}{}^{ab}=\partial_{\mu}\go_{\nu}{}^{ab}
     +\go_{\mu}{}^a{}_c\,\go_{\nu}{}^{cb}
     -2 \gl^2 h_{\mu}{}^a\,h_{\nu}{}^b - (\mu\leftrightarrow \nu) \,,
\ee
\be
\label{tor1}
   R_{\mu\nu}{}^a=\partial_{\mu}h_{\nu}{}^a
     +\go_{\mu}{}^a{}_c h_{\nu}{}^c - (\mu\leftrightarrow \nu) \,,
\ee
which differs from (\ref{rim}), (\ref{tor}) by the terms proportional
to $\gl^2$ on the r.h.s. of (\ref{rim1}).
Again, one can express Lorentz connection $\go_{\mu}{}^{ab}$
via vielbein $h_{\mu}{}^a$ with the aid of the constraint $R_{\mu\nu}{}^a=0$
(for the non-degenerate $h_{\mu}{}^a$).
Substituting $\go_{\mu}{}^{ab} = \go_{\mu}{}^{ab}(h)$ into (\ref{rim1}),
one can see that the condition $R_{\mu\nu}{}^{ab}=0$ is equivalent to
\be
\label{cR}
    {\cal R}_{\mu\nu}{}^{ab}=
     2 \gl^2 (h_{\mu}{}^a\,h_{\nu}{}^b-h_{\nu}{}^a\,h_{\mu}{}^b)
\ee
and therefore describes the AdS space-time
with radius $(\sqrt{2}\gl)^{-1}$.
(${\cal R}\equiv {\cal R}_{\nu\mu}{}^{\mu\nu}= -2 d(d-1) \gl^2 $.)
Note that (\ref{cR}) admits solutions both with the regular
metrics (hyperboloid) and with the singular ones (black holes)
\cite{Hawk}. The analysis below is applicable in the both cases.

Such a formulation of space-time geometry provides a natural starting point
for the formulation of dynamics of free matter fields in terms of
covariant constancy conditions for appropriate (infinite-dimensional)
representations of a chosen space-time symmetry algebra.
Namely, this ``unfolded formulation" \cite{Unf} allows one to rewrite
free field equations in the form
\be
\label{dC}
    dC_i =A_i{}^j \wedge C_j  \,,
\ee
where $d=dx^\nu \frac{\partial}{\partial x^\nu} $ is
the standard space-time exterior
differential, with the gauge fields $A_i{}^j =A^a(T_a)_i{}^j$ obeying the
zero-curvature conditions
\be
\label{dA}
   dA^a=U^a_{bc}\,A^b \wedge A^c \,.
\ee
Here 1-forms $ A^a$ are the gauge fields of $l$
or some its extension $g$ (e.g. supersymmetric or HS ones).
$C_i$ are some $p$-forms which take their values in a certain representation
space of $g$. The zero-curvature condition (\ref{dA}) is a counterpart of
the equations (\ref{zc}). The fact that
$(T_a)_i{}^j$ form some representation $V$ of $l$
guarantees that the compatibility conditions
for (\ref{dC}) are satisfied. For the simplest cases of matter
fields, $C_i$ are 0-forms (cf. the example of a scalar field below).
As we will demonstrate, a natural framework for the relevant representations
is provided by the Moyal bracket.

Thus, the free field problem consists in finding an appropriate
representation of $l$ that leads to the correct field equations.
After the equations are rewritten in the ``unfolded form'',
one can write down their general solution in the pure gauge form
$A(x)=-g^{-1}(x) dg(x)$, $C(x)=T(g^{-1})(x) C_0$, where  $C_0$ is
an arbitrary $x$ - independent element of the representation space of $V$.
This general solution has a structure of the covariantized Taylor
type expansion (see \cite{Unf} for more details).
In the subsequent sections we give a simple realization of
the relevant infinite-dimensional
representation of the $d3$ AdS algebra $o(2,2)$.

At the non-linear level, the problem consists in the construction
of a non-trivial (non-linear in $C$) deformation of the equations
(\ref{dC}) and (\ref{dA}) which preserves their formal consistency.
This is our main concern for the $d3$ problem
considered in this paper. We will follow a method developed
previously in \cite{more,Eq}, which reduces to imposing non-linear
constraints in a larger system of the form
(\ref{dC}), (\ref{dA}) and is in many respects analogous to
the method of Hamiltonian reduction.

Now let us illustrate how this approach works at the linearized level
by the example of the Klein-Gordon equation $(\Box+M^2)\phi=0$ for
a massive scalar field $\phi$ in the $d$-dimensional
flat space-time. We describe
the flat space-time geometry by the zero-curvature
gauge fields (\ref{Poinc}) of the Poincar\'e algebra $iso(d-1,1)$.
To describe dynamics of the spin-0
massive field $\phi(x)$ let us introduce an infinite collection of 0-forms
$\phi_{a_1\ldots a_n}(x)$ $(n=0,1,\ldots )$ that are totally symmetric
tensors obeying the constraints
\be
\label{cons}
   \eta^{bc}\phi_{bca_1\ldots a_n}(x)=-M^2\,\phi_{a_1\ldots a_n}(x) \,,
\ee
where $M$ is an arbitrary constant. Then the unfolded formulation of
the Klein-Gordon equation is
provided by the following infinite chain of equations,
\be
\label{ch}
   D_\nu\phi_{a_1\ldots a_n}(x)=h_\nu{}^b\phi_{ba_1\ldots a_n}(x) \,.
\ee
Here $D_\nu \phi_a = \partial_\nu \phi_a + \go_{\nu a}{}^b \phi_b$ is
the Lorentz-covariant derivative.
As we are in the flat space, one can choose
the gauge (\ref{gk}) and use the ordinary flat derivative $\partial_\nu$
instead of $D_\nu$.
One can easily check that the system (\ref{ch}) is formally consistent
(i.e. compatible with $D^2=0$). Differentiation of the constraints
(\ref{cons}) shows that they are consistent with (\ref{ch})
\footnote{Moreover, all these constraints
for $n>0$ are consequences of (\ref{cons}) and (\ref{ch}) at $n=0$.}.
  These properties are equivalent to the fact that the set of zero-forms
$\phi_{a_1\ldots a_n}(x)$ subject to (\ref{cons})
spans some representation of the Poincar\'e algebra.

To show that the system (\ref{ch}) with the constraints (\ref{cons})
is equivalent to the Klein-Gordon equation with the mass $M$ let us
identify the scalar field $\phi(x)$ with the $n=0$ component of the
tower of tensors $\phi_{a_1\ldots a_n}(x)$. The first two equations
in (\ref{ch}) read
\be
   D_\mu\phi=\phi_{\mu}\,,\quad
        D_\mu\phi_\nu=\phi_{\mu\nu} \,.
\ee
One solves for  $\phi_{\nu}$ from these equations and obtains
\be
\label{fi}
   D_{\mu}D_{\nu}\phi=\phi_{\mu\nu} \,.
\ee
Contracting the indices $\mu\,,\nu$ with the aid
of the Minkowski metrics $\eta^{\mu\nu}$ and taking into account
(\ref{cons}), we get
$$
   (\Box+M^2)\phi=0 \,.         
$$
All the rest equations in (\ref{ch}) express all highest
tensors via higher-order derivatives of $\phi(x)$,
\be
\label{via}
   \phi_{\mu_1\ldots \mu_n}=D_{\mu_1}\ldots D_{\mu_n}\phi \,,
\ee
imposing no additional conditions on the dynamical field $\phi(x)$.
Thus, we see that the Klein-Gordon equation results from the constraints
(\ref{cons}). Note  that without constraints (\ref{cons})
the system (\ref{ch}) is equivalent to (\ref{via}) and therefore
is dynamically empty. This simple example
demonstrates how constraints can impose dynamics.

One can extend this approach to the case of AdS space-time.
In this paper, we consider only the case $d=3$ and use the formalism
of two-component spinors.
We describe the 2+1 dimensional AdS space in terms of
the Lorentz connection one-form
$\go^{\ga\gb}=dx^\nu \go_\nu{}^{\ga\gb}(x)$
and dreibein one-form
$h^{\ga\gb}=dx^\nu h_\nu{}^{\ga\gb} (x)$.
Here $x^\nu$ are space-time coordinates $(\nu =0,1,2)$ and
$\ga,\gb,\ldots =1,2$ are spinor indices which are
raised and lowered with the aid of the symplectic form
$\gep_{\ga\gb}=-\gep_{\gb\ga}$,
$A^{\ga}=\gep^{\ga\gb}A_{\gb}$,
$A_{\ga}=A^{\gb}\gep_{\gb\ga}$,
$\gep_{12}=\gep^{12}=1$. The one-forms
$\go_{\ga\gb}$ and $h_{\ga\gb}$ are symmetric in $\ga$ and $\gb$.
The AdS geometry is described by the equations
\be
\label{d omega}
    d\go_{\ga\gb}=\go_{\ga\gga}\wedge\go_\gb{}^\gga+
\gl^2h_{\ga\gga}\wedge h_\gb{}^\gga   \,,
\ee
\be
\label{dh}
    dh_{\ga\gb}=\go_{\ga\gga}\wedge h_\gb{}^\gga+
  \go_{\gb\gga}\wedge h_\ga{}^\gga \,,
\ee
which have a form of zero-curvature conditions
\footnote{{}From now on we will use a form of equation that differs from
   (\ref{rim1}), (\ref{tor1}) by  $\go \to - \go$.}
for $l = o(2,2)\sim sp(2)\oplus sp(2)$.
For the space-time geometric interpretation of these equations
one has to assume that the dreibein $h_\nu{}^{\ga\gb}$
is a non-degenerate $3\times 3$ matrix, so that
the inverse dreibein $h^\nu{}_{\ga\gb}$ can be defined \cite{Unf},
\be
\label{inv d}
   h_\nu{}^{\ga\gb}h^\nu{}_{\gga\gd}=
    \frac12\,(\gd^\ga_\gga\gd^\gb_\gd+\gd^\ga_
    \gd\gd^\gb_\gga)\,.
\ee
Then, (\ref{dh}) reduces to the zero-torsion condition which expresses
Lorentz connection $\go_\nu{}^{\ga\gb}$ via dreibein
$h_\nu{}^{\ga\gb}$, and (\ref{d omega})
implies that the Riemann tensor 2-form ${\cal R}_{\ga\gb}$
satisfies the AdS vacuum condition
${\cal R}_{\ga\gb}= -\gl^2h_{\ga\gga}\wedge h_\gb{}^\gga$.

The massive Klein-Gordon and Dirac equations in the AdS space-time
take the form
\be
\label{K-G D}
  \Box C=\left(\frac32\gl^2-M^2\right)C \qquad   
   \mbox{and} \qquad
     h^\nu{}_\ga{}^\gb D_{\nu}C_\gb=\frac M{\sqrt2} C_\ga
\ee
for the spin-0 boson and spin-$\frac12$ fermion fields $C(x)$
and $C_{\ga}(x)$. Here $\Box =D^{\mu}D_{\mu}$,
where $D_{\mu}$ is the full background covariant derivative
containing the symmetric Christoffel connection defined
via the metric postulate $D_{\mu}h_{\nu}^{\ga\gb}=0$.
The world indices $\mu$, $\nu$ are raised and lowered by the metric tensor
$g_{\mu\nu}=h_\mu{}^{\ga\gb}h_\nu{}_{\ga\gb}$.
  To reformulate (\ref{K-G D}) in a form of covariant
constancy conditions (\ref{dC}) we introduce an infinite set of the
symmetric multispinors $C_{\ga_1\ldots\ga_n}$ for all $n\ge 0$.
({}From now on we will
assume total symmetrization of the indices denoted by the same letter
and will often use the notation
$C_{\ga(n)}=C_{\ga_1 \dots \ga_n}$ when only
a number of indices is important.)
As shown in \cite{Unf}, the following two infinite chains of
equations,
\be
\label{chainbos}
   DC_{\ga(n)} = h^{\gb\gga}C_{\gb\gga\ga(n)}+
        n(n-1)\left( {\gl^2\over 4}-{M^2\over 2(n^2-1)} \right)
        h_{\ga\ga}C_{\ga(n-2)}
\ee
for even $n$, and
\be
\label{chainferm}
  DC_{\ga(n)} = h^{\gb\gga}C_{\gb\gga\ga(n)} -
     \frac {\sqrt{2} M}{n+2}\; h_{\ga}{}^{\gb}C_{\gb\ga(n-1)}
     + n(n-1)\left( {\gl^2\over 4}-{M^2\over 2n^2} \right)
          h_{\ga\ga}C_{\ga(n-2)}
\ee
for odd $n$, where $D$ is the background Lorentz-covariant differential,
\be
\label{Lcov D0}
    D C_{\ga (n)}=dC_{\ga (n)}+n\go_{\ga}{}^\gga
       C_{\gga\ga(n-1)} \,,
\ee
are equivalent respectively to the
Klein-Gordon and Dirac equations (\ref{K-G D}) for the lowest components
$C$ and $C_\ga$. Similarly to the flat space example considered before,
all the rest equations contained in
(\ref{chainbos}) and (\ref{chainferm}) either  express
highest multispinors via highest space-time derivatives of $C$ or
$C_{\ga}$, or reduce to some identities.

\section{Operator Realization for Arbitrary Mass}\label{OpReal}

Let us now describe an algebraic construction that leads automatically
to the correct massive field equations \cite{BPV} in $d=3$.
Following to \cite{OP2}, we introduce ``oscillators''
$\hat{y}_\ga$ $(\ga,\gb =1,2)$ obeying the commutation relations
\be
\label{y mod}
  [\hat{y}_\ga,\hat{y}_\gb]=2i\gep_{\ga\gb}(1+\nu k) \,,\qquad
  k\hat{y}_\ga=-\hat{y}_\ga k \,, \qquad  k^2 =1 \,,
\ee
where $[a,b] = ab-ba$, and $\nu$ is a free parameter.
The operators $\hat{y}_\ga$ and $k$ are treated here as generating elements
of the associative algebra $Aq(2,\nu)$
(in the notation of \cite{OP2}), the enveloping algebra of the relations
(\ref{y mod}).
We call $k$ Klein operator.

The main property of these oscillators is that the bilinears
\be
\label{T}
   T_{\ga\gb} =\frac{1}{4i} \{\hat{y}_\ga ,\hat{y}_\gb\}
\ee
$(\{a,b\} = ab+ba)$  fulfill the standard $sp(2)$ commutation relations
\be
\label{sp(2) com}
  [T_{\ga\gb},T_{\gga\gd}]=
    \gep_{\ga\gga}T_{\gb\gd}+
    \gep_{\gb\gd}T_{\ga\gga}+
    \gep_{\ga\gd}T_{\gb\gga}+
    \gep_{\gb\gga}T_{\ga\gd} \,,
\ee
and
\be
\label{oscom}
  [T_{\ga\gb} ,\hat{y}_{\gga}]=
    \gep_{\ga\gga}\hat{y}_{\gb}+
    \gep_{\gb\gga}\hat{y}_{\ga} \,
\ee
for any $\nu$. A specific realization of these oscillators
was originally considered by Wigner
\cite{Wig} who addressed a question whether it is possible to
modify the oscillator commutation relations in such a way that
the relation $[H, a^\pm ]=\pm a^\pm $ remains valid.
The latter is a particular case of (\ref{oscom}) with
$H=T_{12}$ and $a^\pm =y_{1,2}$.

The property (\ref{sp(2) com}) allows us to realize
the $o(2,2)$ gravitational connection forms as
\be
\label{W gr}
  W_{gr} (x)= \go +\gl\psi h \,;\qquad
    \go=\frac1{8i}\,\go^{\ga\gb}\{\hat{y}_\ga,
    \hat{y}_\gb\} \, ,
    \quad h=\frac1{8i}\,h^{\ga\gb}\{\hat{y}_\ga,\hat{y}_\gb\} \,,
\ee
where $\psi$ is an additional central involutive element,
\be
\label{psi}
   \psi^2=1\,,\qquad [\psi,\hat{y}_{\ga}]=0\,,\qquad
   [\psi,k]=0\,,
\ee
which is introduced to describe the doubling of $sp(2)$ in
the $d3$ AdS algebra $o(2,2)\sim sp(2)\oplus sp(2)$.
The generators for the Lorentz rotations (diagonal $sp(2)$)
and AdS translations are
\be
\label{al}
   L_{\ga\gb}=\frac1{4i}\{\hat{y}_\ga,
    \hat{y}_\gb\}\,,\qquad \,
   P_{\ga\gb}=\frac1{4i}\{\hat{y}_\ga,
    \hat{y}_\gb\}\psi  \,,
\ee
while the components in the direct sum $sp(2)\oplus sp(2)$ are spanned
by the combinations $L_{\ga\gb}\pm P_{\ga\gb}$ projected out by
${1\over 2}(1\pm\psi)$.
Now the equations (\ref{d omega}) and (\ref{dh}), which
describe the AdS geometry, read
\be
\label{va}
   dW_{gr} =W_{gr} \wedge W_{gr}\, .
\ee
Because of the property (\ref{sp(2) com})
the $\nu$-dependence does not appear explicitly in (\ref{va}).

Let us introduce following to
\cite{BPV} the operator-valued 0-form
\be
\label{hatC}
  C(\hat{y},k,\psi|x)=\sum_{A,B=0,1}
  \sum_{n=0}^\infty \frac 1{n!} \gl^{-[\frac n2]}
  C^{AB}_{\ga_1 \ldots\ga_n}(x) k^A
  \psi^B\hat{y}^{\ga_1}\ldots \hat{y}^{\ga_n}\, ,
\ee
where $C^{AB}_{\ga_1 \ldots\ga_n}$ are totally symmetric multispinors
(this implies the totally symmetric
(Weyl) ordering with respect to $\hat{y}_{\ga}$).
The following two types of equations,
\be
\label{aux}
    DC^{aux}=\gl\psi\, [h,C^{aux}] \,
\ee
and
\be
\label{dyn}
    DC^{dyn}=\gl\psi\, \{h,C^{dyn}\} \, ,
\ee
where
\be
\label{dif}
    DC=dC-[\go,C] \,,
\ee
are formally consistent, i.e. the integrability conditions are satisfied
as a consequence of the vacuum conditions (\ref{va}). Indeed,
(\ref{aux}) corresponds to the adjoint action of the
space-time algebra $o(2,2)$ (\ref{al}) on the algebra of modified
oscillators. The equation (\ref{dyn}) corresponds to another
representation of the space-time symmetry which we call twisted
representation. The fact that one can replace commutator
by anticommutator in the $h$-dependent term
without destroying the formal consistency is a
consequence of the property that the AdS algebra possesses an involutive
automorphism that changes a sign of the AdS translations.
In the particular realization we use it is induced by the
automorphism $\psi \to - \psi$.

There is an important difference between these two representations.
The one corresponding to (\ref{aux}) decomposes
into an infinite direct sum of finite-dimensional representations of
$o(2,2)$. Because of the property
(\ref{oscom}) this representation is $\nu$-independent and, therefore,
eq.~(\ref{aux}) at any $\nu$ is equivalent to that with $\nu=0$.
The latter was shown in \cite{Unf} to describe an infinite set of
auxiliary (topological) fields,
each carrying no dynamical degrees of freedom
(in a topologically trivial situation). On the other hand,
the twisted representation is just the infinite-dimensional representation
needed for the description of the matter fields \cite{BPV}
(to simplify notations we will sometimes
use the symbol $C$ for the twisted representation $C^{dyn}$).

To see this one has to carry out a component analysis
of the equations (\ref{dyn}), which consists of some operator
reorderings bringing all terms into the Weyl-ordered form
with respect to $\hat{y}_\ga$.
It is convenient to replace all operators by their Weyl symbols, so that
$C(\hat{y},k,\psi)\rightarrow C(y,k,\psi)$ according to the rule
\be
\label{Wr}
  C^{AB}_{\ga_1 \ldots\ga_n}(x) k^A
  \psi^B\hat{y}^{\ga_1}\ldots \hat{y}^{\ga_n} \rightarrow
  C^{AB}_{\ga_1 \ldots\ga_n}(x) k^A
  \psi^B y^{\ga_1}\ldots y^{\ga_n}   \,,
\ee
with
\be
\label{y_sym}
  [y_\ga,y_\gb]=0 \,,\quad ky_\ga=-y_\ga k\,,\quad k^2 =1 \,,
     \quad \psi^2=1 \,,\quad [\psi, k]=0 \,,\quad
     [\psi, y_\ga]=0 \,.
\ee
As a result, (\ref{dyn}) takes the form
\be
\label{DC mod}
  DC(y,k,\psi)=\psi h^{\ga\gb} \left[a(N)
  \frac{\partial}{\partial y^\ga }\frac{\partial}{\partial y^\gb}+
  b(N) y_\ga\frac{\partial}{\partial y^\gb}+ e(N)
  y_\ga y_\gb \right]C(y,k,\psi) \,,
\ee
where $D$ is the Lorentz-covariant differential,
$ D=d-\go^{\ga\gb}y_\ga \frac{\partial}{\partial y^\gb}$,
and $N$ is the Euler operator,
$N=y^\ga\frac{\partial}{\partial y^\ga}$.
The coefficients $a(n)$, $b(n)$ and $e(n)$ are \cite{BPV}
\bee
   a(n) &=& \frac{i\gl}2 \left[1+\nu k\frac{1+(-1)^n}{(n+2)^2-1}
  -\frac{\nu^2}{(n+2)^2((n+2)^2-1)} \left((n+2)^2-
   \frac{1-(-1)^n}2 \right)\right] \,,   \nonumber  \\
   && {}  \label{a} \\
\label{b}
  b(n) &=& -\nu k\gl\,\frac{1-(-1)^n}{2n(n+2)}\,,\quad n>0,
    \qquad b(0)=0 \,,   \\
\label{e}
  e(n) &=& -\frac{i\gl}2  \,.
\eee

Due to the presence of the Klein operator
$k$ we have a doubled number of fields compared to the analysis
of \cite{Unf}. The invariant subsets
can be projected out
with the aid of the projectors $P_\pm$,
\be
\label{pm}
   C^\pm=P_\pm C\, ,\qquad P_\pm=\frac{1\pm k}2\, .
\ee
Then, (\ref{DC mod}) leads to
the following two infinite chains of equations
for bosons and fermions,
\be
\label{chainbos+-}
   DC^{\pm}_{\ga(n)}=\frac i2\psi\left[\left(1-
    \frac{\nu(\nu\mp2)}{(n+1)(n+3)}
   \right) h^{\gb\gga}C^{\pm}_{\gb\gga\ga(n)}-
   \gl^2n(n-1)h_{\ga\ga}C^{\pm}_{\ga(n-2)}\right]
\ee
for even $n$, and
\begin{eqnarray}
\label{chainferm+-}
  DC^{\pm}_{\ga(n)} & = & \frac i2\psi\left(1-\frac{\nu^2}{(n+2)^2}\right)
     h^{\gb\gga}C^{\pm}_{\gb\gga\ga(n)} \pm
     \psi\frac {\nu\gl}{n+2}h_{\ga}{}^{\gb}C^{\pm}_{\gb\ga(n-1)}
     \nonumber\\
   & & {}-\frac i2 \psi\gl^2n(n-1)h_{\ga\ga}C^{\pm}_{\ga(n-2)}
\end{eqnarray}
for odd $n$ ($DC^{\pm}_{\ga(n)}$ is given in (\ref{Lcov D0})).
These chains of equations differ from (\ref{chainbos}) and (\ref{chainferm})
by some rescalings of the fields $C_{\ga(n)}$ and
lead \cite{BPV} to the massive field equations (\ref{K-G D})
for the lowest rank components $C$ and $C_{\ga}$.
The values of mass are related to the parameter $\nu$ as follows,
\be
\label{M}
  M^2_\pm =\gl^2\frac{\nu(\nu\mp 2)}2 \,
\ee
for bosons and
\be
\label{M f}
  M^2_\pm =\gl^2\frac{\nu^2}2 \,
\ee
for fermions. The signs ``$\pm$'' are in agreement with those in
(\ref{pm}).

The construction above generalizes in a natural way the realization
of the equations for massless matter fields in terms of
the ordinary ($\nu=0$) oscillators proposed in \cite{Unf}.
An important comment however is that
for arbitrary $\nu$ this construction
not necessarily leads to non-vanishing coefficients $a(n)$
in (\ref{DC mod}). Consider for instance the equation (\ref{chainbos+-})
for $C^+$ (i.e. set $k=1\,,\,n=2m$, $m\in {\bf Z}$).
We observe that some of the coefficients in front of the first term on
the right hand side of (\ref{chainbos+-})
vanish if $\nu=2l+1$, $\forall l\in {\bf Z}$.
This conclusion is in agreement with the results
of \cite{OP2}, where it was shown that for these values of $\nu$ the
algebra $Aq(2,\nu )$ possesses ideals.
Thus, at $\nu =2l+1$ some of the
rescalings of fields mentioned above degenerate
and the system of equations derived from
the operator realization of (\ref{dyn})
turns out to be different from that considered in
\cite{Unf}. It is important to stress that this does not imply any
essential singularity. Rather, for $\nu =2l+1$ the chains of equations
start not from a scalar but from some multispinor component.
Interesting enough, the cases of $\nu = 1,3$ correspond to the
massless electrodynamics \cite{BPV}.
The specificities of the degenerated systems with $\nu=2l+1$
is a very interesting issue which needs an independent study
and will be considered elsewhere. In this paper,
we focus on the generic case of the non-degenerate $\nu$.

\section{Star Product}\label{StarPr}

The generating functions introduced in the next section to
describe non-linear $d=3$ HS dynamics will be treated as
elements of the associative algebra with the following product law,
\be
\label{prod}
   (f*g)(z,y)=\frac{1}{(2\pi)^2}\int d^2ud^2v\;\exp(iu_\ga v^\ga)
   f(z+u,y+u)g(z-v,y+v) \,,
\ee
where $z_\ga, y_\ga, u_\ga$ and $ v_\ga$ are spinor variables.
It is easy to see that this product law yields a particular
realization of the Heisenberg-Weyl algebra with the defining relations
\be
\label{y,z}
   [y_\ga,y_\gb]_*=-[z_\ga,z_\gb]_*=2i\gep_{\ga\gb}\,,\quad
   [y_\ga,z_\gb]_*=0 \,
\ee
($[a,b]_*=a*b-b*a$). These commutation relations are particular
cases of the following simple formulae
\be
\label{y,f}
    [y_{\ga}, f]_*=2i{\partial f\over \partial y^\ga}  \,,
\ee
\be
\label{z,f}
    [z_{\ga}, f]_*=-2i{\partial f\over \partial z^\ga} \,,
\ee
which are true for an arbitrary $f(z,y)$ (e.g. from the
class defined in Appendix A).

With the aid of (\ref{prod}) one can check that the operator
$\exp(iz_\ga y^\ga)$
possesses the properties of the inner Klein operator,
\be
\label{e^}
   e^{i(z_\ga y^\ga)} * f(z,y) = f(-z,-y) * e^{i(z_\ga y^\ga)}
\ee
for an arbitrary function $f(z,y)$, and
\be
   e^{i(z_\ga y^\ga)} * e^{i(z_\gb y^\gb)} = 1\,.
\ee

The star product (\ref{prod}) corresponds
to the normal ordering of the Heisenberg-Weyl
algebra with respect to the generating elements
\be
   a^+_\ga = \frac{1}{2} (y_\ga - z_\ga )\,,\qquad
   a_\ga = \frac{1}{2} (y_\ga + z_\ga )  \,,
\ee
which satisfy the commutation relations
\be
\label{com a}
   [a_\ga, a_\gb]_*=[a^+_\ga, a^+_\gb] =0 \,,\quad
   [a_\ga, a^+_\gb]_* =i\gep_{\ga\gb}
\ee
and are interpreted as creation and annihilation operators. This is
most evident from the relations
\be
\label{pr a}
   a^+_\ga * f(a^+, a) =
      a^+_\ga f(a^+, a) \,,\qquad
   f(a^+, a) * a_\ga =
      f(a^+, a)  a_\ga \,.
\ee

{}From (\ref{com a}) and (\ref{pr a}) it follows that
\be
\label{pr af}
   a_\ga f(a^+_\gga a^\gga)\; * \; a^\ga f(a^+_\gga a^\gga) = 0
    \,,\qquad
   a^+_\ga f(a^+_\gga a^\gga)\; * \;
      a^{+\ga} f(a^+_\gga a^\gga) = 0 \,,
\ee
because
$a_{\ga} a^{\ga} = a^{+}_{\ga} a^{+\ga} \equiv 0$.

The  formula (\ref{prod}) is an integral version
\cite{bershub} of the Moyal star product \cite{MOY1, MOY2}. The standard
differential formulae for the Moyal product
can be obtained from (\ref{prod}) by the Gaussian integration.
The integral formula turns out to be more convenient in many respects.

To formulate the full nonlinear system that possesses
all necessary gauge symmetries we will need an
associative algebra ${\cal A}$ that is generated by the commuting
spinors $z_\ga$, $y_\ga$ $(\ga=1,2)$,
\be
    [y_\ga,y_\gb]=[z_\ga,z_\gb]=[z_\ga,y_\gb]=0\,,
\ee
a pair of Clifford elements $\psi_i\: (i=1,2)$,
$\{\psi_i,\psi_j\}=2\gd_{ij}$
that commute to all other generating elements,
and another pair of Clifford-type
elements $k$ and $\rho$ that have the following properties,
\be
\label{Klein}
    k^2=1\,,\: \rho^2 =1 \,,\: k\rho =-\rho k \,,\:
    ky_\ga=-y_\ga k\,,\: kz_\ga=-z_\ga k\,, \:
    \rho y_\ga=y_\ga \rho \,,\: \rho z_\ga=z_\ga \rho \,.
\ee
Thus, a generic element of ${\cal A}$ has a form
\bee
\label{gexp}
   A(z,y;\psi_{1,2},k,\rho) & = & \sum_{B,C,D,E=0}^1 \;
     \sum_{m,n=0}^\infty \;\frac 1{m!n!} \;
     A^{BCDE}_{{\ga_1}\ldots {\ga_m}{\gb_1}\ldots {\gb_n}} \;
     k^B \rho^C \psi_1^D \psi_2^E \nonumber \\
     & & {} \times  z^{\ga_1}\ldots z^{\ga_m}\, y^{\gb_1}\ldots y^{\gb_n} \,.
\eee

In the sector of spinor variables the product law is defined
according to (\ref{prod}).
Since the elements $\psi_1\,,\,\psi_2\,,\,k\,,\,\rho$
do not commute, the order of the generating elements is essential.
Namely, consider some monomials $f=F(z,y)\Phi(\psi_{1,2},k,\rho )$ and
$g=G(z,y)\Psi(\psi_{1,2},k,\rho )$.
Due to the sign changes in (\ref{Klein}) we have
\be
    \Phi(\psi_{1,2},k,\rho ) G(z,y) = \tilde{G} (z,y)
    \Phi(\psi_{1,2},k,\rho )
\ee
with some $\tilde{G}$. The full product is then defined as
\be
   (f*g)(z,y;\psi_{1,2},k,\rho) = (F*\tilde{G}) (z,y) \;
   \Phi(\psi_{1,2},k,\rho ) \Psi(\psi_{1,2},k,\rho )\,.
\ee

Let us remind the reader that invertible mappings
$\dagger$ and $\gs$ of
some algebra ${\cal F}$ to itself that have the properties
\be
\label{pr inv1}
    (\gl_1 \, a_1+\gl_2 \, a_2)^\dagger =
       \bar\gl_1\, a_1^\dagger + \bar\gl_2\, a_2^\dagger \,,
\ee
\be
\label{pr inv2}
    (a_1 * a_2)^\dagger = a_2^\dagger * a_1^\dagger \,,\qquad
    (a^\dagger)^\dagger = a \,,
\ee
and
\be
\label{pr1}
    \gs(\gl_1 \, a_1 + \gl_2 \, a_2)=
       \gl_1\, \gs(a_1) + \gl_2\, \gs(a_2) \,,\qquad
\ee
\be
\label{pr2}
    \gs(a_1 * a_2) = \gs(a_2) * \gs(a_1) \,,\qquad
    \forall a, a_1, a_2 \in {\cal F}\,, \qquad
    \forall \gl_1, \gl_2 \in {\bf C}    \,,
\ee
are called involution and antiautomorphism, respectively.
The algebra ${\cal A}$ admits the following important involution
$\dagger$ and antiautomorphism $\gs$ \cite{Ann, KV1},
\be
\label{inv}
    [A(z,y;\psi_{1,2},k,\rho)]^\dagger =
           \bar{A}^{rev}(-z,y;\psi_{1,2},k,\rho) \,,
\ee
\be
\label{sigma}
    \gs [A(z,y;\psi_{1,2},k,\rho)] =
            A^{rev}(-iz,iy;\psi_{1,2},k,\rho) \,,
\ee
where the notation $A^{rev}(...)$ means that an order of all generating
elements in the monomial expressions on the r.h.s. of (\ref{gexp})
is reversed. The properties (\ref{pr inv1})-(\ref{pr2}) can be easily
checked with the aid of (\ref{prod}).
The important property of the transformations
(\ref{inv}) and (\ref{sigma}) is that the inner Klein element
$\exp(iz_\ga y^\ga)$ and the element $K$,
\be
\label{K}
    K=k e^{i(zy)}\,,\qquad (zy)=z_\ga y^\ga \,,
\ee
are invariant,
\be
    (e^{i(zy)})^\dagger = e^{i(zy)} \,,\qquad K^\dagger = K \,,
\ee
\be
    \gs(e^{i(zy)}) = e^{i(zy)} \,,\qquad \gs(K) = K \,.
\ee

\section{Nonlinear System with Arbitrary Mass}\label{NonlSys}

To formulate the full nonlinear system that possesses
all necessary gauge symmetries and reduces at the linearized level
to the free system described in sect.~\ref{OpReal}
we introduce three types of the generating functions
$W$, $B$, and $S_\ga$ that take values in ${\cal A}$ and
depend on the space-time coordinates $x^\nu$  $(\nu=0,1,2)$.
$W$ is a space-time 1-form, $W=dx^\nu W_\nu(z,y;\psi_{1,2},k,\rho | x)$,
while $B = B(z,y;\psi_{1,2},k,\rho | x)$ and
$S_\ga = S_\ga(z,y;\psi_{1,2},k,\rho | x)$ are space-time 0-forms.

We start with the system of equations analogous to the $d4$
system of \cite{more} and $d3$ massless system of \cite{Eq},
\be
\label{WW}
      dW=W*\wedge W \,,
\ee
\be
\label{WB}
      dB=W*B-B*W   \,,
\ee
\be
\label{WS}
      dS_\ga=W*S_\ga-S_\ga*W   \,,
\ee
\be
\label{SS}
      S_\ga * S^\ga=-2i(1+B*K)   \,,
\ee
\be
\label{SB}
      S_\ga*B=B*S_\ga \,,
\ee
where $K$ is given by (\ref{K}).

The multispinorial coefficients
$A^{BCDE}_{{\ga_1}\ldots {\ga_m}{\gb_1}\ldots {\gb_n}}$
in the expansions (\ref{gexp}) of the functions
$A(z,y;\psi_{1,2},k,\rho | x) = W_\mu$, $B$, or $S_\ga$
carry standard Grassmann parity $\pi=0$ or $1$ in accordance with the
number of spinor indices,
\be
\label{pWB}
   \pi(W_{\mu,\,\,\ga(m)\gb(n)})=|m+n|_2 \,,\quad
   \pi(B_{\ga(m)\gb(n)})=|m+n|_2 \,,
\ee
\be
\label{pS}
   \pi(S_{\gga,\,\,\ga(m)\gb(n)})=|m+n+1|_2 \,,
\ee
where
$$
\begin{array}{rcl}
   |n|_2 & = & \left\{ \begin{array}{rcl}
                          0\,, & n\mbox{ - even} \\
                          1\,, & n\mbox{ - odd} \,,
                       \end{array}
               \right.
\end{array}
$$
and are defined to commute with the generating elements
$z_\ga\,,\,y_\ga\,,\,k\,,\,\rho\,,\,\psi_{1,2}$.
As a result, the commutators on the
right hand sides in (\ref{WW})-(\ref{WS})
 reduce to supercommutators in terms of
the polynomials of $z_\ga\,,\,y_\ga\,,\,k\,,\,\rho$ and $\psi_{1,2}$.

With the aid of the obvious involutive automorphism $\rho\to -\rho$
of ${\cal A}$, combined with the symmetry of the equations
$S_\ga\to -S_\ga$, we truncate the system (\ref{WW})-(\ref{SB})
to the one with $W$ and $B$ independent of $\rho$
and $S_\ga$ linear in $\rho$,
\be
\label{tru}
    \!\!\! W(z,y;\psi_{1,2},k,\rho | x)=W(z,y;\psi_{1,2},k | x)\,,\quad
    B(z,y;\psi_{1,2},k,\rho | x)=B(z,y;\psi_{1,2},k | x)\,,
\ee
\be
\label{truS}
    S_\ga(z,y;\psi_{1,2},k,\rho | x)=
      \rho s_\ga(z,y;\psi_{1,2},k | x)\,.
\ee
It is this reduced system which is discussed below
and will be shown to describe the dynamics of massive matter fields.
Now the gauge fields $W$ and the matter fields $B$ take values in the
algebra $A$, the subalgebra of ${\cal A}$ spanned by
$\rho$-independent elements.

For this system one finds taking into account (\ref{e^}), (\ref{Klein}),
and (\ref{K}) that
\be
\label{propK}
    K*W=W*K\,,\qquad K*B=B*K\,,\qquad
\ee
and
\be
\label{propK1}
    K*S_{\ga}=-S_{\ga}*K \,.
\ee
The additional minus sign in
(\ref{propK1}) is due to the factor of $\rho$ in (\ref{truS}).
The dynamical system we study in this paper
differs from that proposed in \cite{Eq} by
the redefinitions $B\to -iB*K\,,\,S_\ga \to \rho S_{\ga}$
and by the doubling of the fields due to the extra dependence
on $k$ ($k^2=1$). As we shall see below, the latter property is
essential for the description of massive supersymmetric matter multiplets.

Let $l$ be the Lie superalgebra constructed from the associative algebra $A$
via (anti)com\-mutators. The equations (\ref{WB}) and (\ref{WS}) describe
0-forms in the adjoint representation of $l$.
Therefore, the system (\ref{WW})-(\ref{SB}) contains zero-curvature
equations (\ref{WW}), covariant constancy conditions (\ref{WB}),
(\ref{WS}) and the constraints (\ref{SS}), (\ref{SB}) which do not
contain space-time derivatives. These constraints make the full
system dynamically non-trivial.

The equations (\ref{WW})-(\ref{SB}) are general coordinate
invariant because of using the exterior algebra formalism
and are invariant under the infinitesimal HS gauge
transformations
\be
\label{delta W}
       \gd W=d\gvep-W*\gvep+\gvep * W \,,
\ee
\be
\label{delta B}
       \gd B=\gvep *B-B*\gvep        \,,
\ee
\be
\label{delta S}
       \gd S_\ga=\gvep *S_\ga-S_\ga*\gvep \,,
\ee
where $\gvep=\gvep(z,y;\psi_{1,2},k|x)$ is an arbitrary
($\rho$-independent) gauge parameter.
The main technical problem now consists in elucidating
a physical content of these equations by proving that they indeed
describe correct relativistic dynamics at the linearized level and
beyond. To show this one should first find an appropriate vacuum
solution of the equations and then analyze perturbations. These problems
are solved in the two subsequent sections.

Before going into technical details, let us show
that the form of the constraints (\ref{SS}), (\ref{SB})
is fixed by the very simple requirement that local Lorentz symmetry
remains undeformed in all orders in interactions
\cite{Prop, Tbilisi, buck1, vol}.

The infinitesimal local Lorentz transformations with the parameter
$\eta^{\ga\gb}(x)$ are
\be
\label{loc L}
   \gd f=[\eta^{\ga\gb}L^{tot}_{\ga\gb}\,,\; f]_* \,,
\ee
where $L^{tot}_{\ga\gb}$ are the generators of the inner
$sp(2)\sim o(1,2)$ automorphism of the algebra,
\be
   L^{tot}_{\ga\gb}={i\over 4}(\{z_{\ga}, z_{\gb}\}_*-
     \{y_{\ga}, y_{\gb}\}_*) \,.
\ee
Actually, with the aid of (\ref{y,f}) and (\ref{z,f}) we see
that these generators rotate properly the
spinor generating elements,
$$
   \gd z_{\ga} = 2\eta_\ga{}^\gb z_{\gb} \,,\qquad
   \gd y_{\ga} = 2\eta_\ga{}^\gb y_{\gb} \,.
$$
Thus, the system (\ref{WW})-(\ref{SB}) is explicitly invariant under local
Lorentz transformations (\ref{loc L}).  However, this symmetry is
spontaneously broken due to the constraint (\ref{SS}) because the r.h.s.
of (\ref{SS}) has a non-vanishing vacuum value and therefore $S_{\ga}$
itself must have a non-vanishing vacuum value.  Therefore, the question
is whether there exists another local Lorentz
symmetry that rotates properly spinor indices of the dynamical
fields and leaves the vacuum solution invariant. Existence of such Lorentz
symmetry in all orders in interactions is a non-trivial property which fixes
a form of the constraints (\ref{SS}), (\ref{SB}).

Indeed, because of (\ref{propK1}) the constraints (\ref{SS}), (\ref{SB})
have a form of the deformed oscillator algebra (\ref{y mod}).
The elements
\be
   M_{\ga\gb}={i\over 4} \{S_\ga, S_\gb\}_*
\ee
are analogous to (\ref{T}) and therefore
obey the Lorentz commutation relations and
rotate properly $S_{\ga}$,
\be
\label{M_}
   [M_{\ga\gb}\,,\;S_{\gga}]_*=\gep_{\ga\gga} S_{\gb}+
       \gep_{\gb\gga} S_{\ga} \,.
\ee
Then, we introduce the generators
\be
\label{l}
   l_{\ga\gb}=L^{tot}_{\ga\gb} - M_{\ga\gb} \,,
\ee
which also satisfy the $sp(2)$ commutation relations
\footnote{This is the standard situation with the diagonal algebra
($L^{tot}_{\ga\gb}$)
of two subalgebras spanned by
$l_{\ga\gb}$ and $M_{\ga\gb}$.}.
  Taking into account (\ref{SB}), we obtain
\be
\label{del B}
   \gd B=[\eta^{\ga\gb} l_{\ga\gb}\,,\; B]_*=
           \eta^{\ga\gb} [L^{tot}_{\ga\gb}\,,\; B]_* \,,
\ee
i.e. $l_{\ga\gb}$ rotate properly the field $B$
which is shown below to describe matter fields.

For the gauge fields $W$ we obtain
\be
\label{del W}
   \gd W = D(\eta^{\ga\gb} l_{\ga\gb})=
       (d\eta^{\ga\gb}) l_{\ga\gb} +
       \eta^{\ga\gb} [L^{tot}_{\ga\gb}\,,\; W]_* \,.
\ee
Here $D(f)=df-[W,f]_*$ and therefore
$D(\eta^{\ga\gb})=d(\eta^{\ga\gb})$, since $\eta^{\ga\gb}(x)$
is independent of any auxiliary variables
(i.e. it is proportional to the unit element of $A$). Also, $D(l_{\ga\gb}) =
[L^{tot}_{\ga\gb}, W]_*$ because $d L^{tot}_{\ga\gb} = 0$ and
$D M_{\ga\gb} = 0$
(cf. eq.(\ref{WS})). {}From (\ref{del W}) one concludes that the gauge
field for a true local Lorentz symmetry is
\be
\label{Wl} W_L = \go_L^{\ga\gb}
l_{\ga\gb} \,,
\ee
while the other gauge fields are rotated properly under the
local Lorentz transformations.

The auxiliary field $S_{\ga}$ is expressed (up to the gauge ambiguity)
via $B$ by the constraint (\ref{SS}) (cf. eq.(\ref{S_1})).
Its transformation law therefore is
\be
\label{}
   [\eta^{\ga\gb} L^{tot}_{\ga\gb}\,,\; S_\gga(B)]_* =
      \eta^{\ga\gb}(\gep_{\ga\gga} S_{\gb} (B)
      + \gep_{\gb\gga} S_{\ga} (B) )
      + \frac{\gd S_\gga}{\gd B}\;\gd B  \,,
\ee
where $ \gd B = \eta^{\ga\gb} [L^{tot}_{\ga\gb}, B]_* $.
Making use of (\ref{M_}), we find
\be
\label{del S}
   \gd S_\gga = [\eta^{\ga\gb} l_{\ga\gb}\,,\; S_\gga]_*=
       \frac{\gd S_\gga}{\gd B}\;\gd B  \,.
\ee
As a result, the $sp(2)$ local Lorentz rotations induced by $l_{\ga\gb}$
do not act on the index $\gga$ of $S_\gga$, acting only on the physical
fields $B$.

Thus, the form of the constraints (\ref{SS}), (\ref{SB}) indeed guarantees
that the local Lorentz symmetry remains unbroken.

\section{Vacuum Solutions}\label{Vacuum}

We consider  vacuum solutions of the system
(\ref{WW})-(\ref{SB}) with
\be
\label{B_0}
   B_0=\nu\,,
\ee
where $\nu$ is some constant independent of the space-time coordinates
and auxiliary variables. As a consequence of (\ref{B_0}),
the vacuum fields $W_0\,,\,S_{0\ga}$ have to satisfy
\be
\label{WW_0}
   dW_0=W_0 *\wedge W_0\,,
\ee
\be
\label{WS_0}
   dS_0=W_0*S_{0\ga}-S_{0\ga}*W_0\,,
\ee
\be
\label{SS_0}
   [S_{0\ga},\,S_{0\gb}]_*=-2i\gep_{\ga\gb}(1+\nu K)  \,,
\ee
while the equations (\ref{WB}) and (\ref{SB}) are trivially satisfied.

Due to (\ref{propK1}) we see that (\ref{SS_0}) has a form of
the deformed oscillator relations (\ref{y mod}).
For $\nu=0$, the standard choice is $S_{0\ga}=\rho z_\ga$.
For general $\nu$, a class of solutions of the equations (\ref{SS_0})
is found in Appendix B. Here we describe the following three
most important solutions,
\be
\label{S_0}
   S_{0\ga}^{\pm}=\rho \left(z_\ga+\nu (z_\ga\pm y_\ga)
    \int_0^1dtt e^{it(zy)} k \right) \,,
\ee
and
\bee
\label{S sym}
   S^{sym}_{0\ga} (z,y) & = & \rho z_\ga - \rho \, {\nu\over 8} \;
     \int^1_{-1} ds (1-s)
    \left[ e^{ \frac{i}{2} (s+1) (zy) }(y_\ga + z_\ga) \; * \;
     \Phi \left({1\over 2}\,,\; 2\,; \; -K \, \mbox{ln} \,|s|^{\nu} \right)
        \right.  \nonumber \\
     & & \left. {} + e^{ \frac{i}{2} (s+1) (zy) }(y_\ga - z_\ga) \; * \;
     \Phi \left({1\over 2}\,,\; 2\,; \; K \, \mbox{ln} \,|s|^{\nu}\right)
        \right] \; * \;K  \,,
\eee
where $\Phi(a,c;x)$ is the degenerate hypergeometric function
(see Appendix B for more detail) and $K$ and $(zy)$ are  defined in (\ref{K}).
The ambiguity in the solutions of the equation (\ref{SS_0})
originates from the gauge transformation (\ref{delta S}).
All three solutions $S^\pm_{0\ga}$ and $S^{sym}_{0\ga}$ belong to
the same gauge equivalence class.
It is easy to see that $S^\pm_{0\ga}$ solve (\ref{SS_0}) by direct
insertion into (\ref{SS_0}) (making use of (\ref{pr af})).
To prove that $S^{sym}_{0\ga}$ solves (\ref{SS_0}) is more tricky
(see Appendix B).

The important properties of
$S^\pm_{0\ga}$ and $S^{sym}_{0\ga}$ are
\be
   S^\pm_{0\ga} (z,y;k,\rho) = - \bar{S}^\mp_{0\ga} (-z,y;k,\rho) \,,\qquad
   S^\pm_{0\ga} (z,y;k,\rho) = i S^\mp_{0\ga} (-iz,iy;k,\rho) \,,
\ee
and
\be
\label{pr Ss}
   S^{sym}_{0\ga} (z,y;k,\rho) = - \bar{S}^{sym}_{0\ga} (-z,y;k,\rho)
    \,,\qquad
   S^{sym}_{0\ga} (z,y;k,\rho) = i S^{sym}_{0\ga} (-iz,iy;k,\rho) \,.
\ee
The solution $S^{sym}_{0\ga}$ is fixed uniquely by the properties
(\ref{pr Ss}). In the part of the analysis independent of the
particular form of a vacuum solution, the symbol $S_{0\ga}$
will be used for any one of the above solutions. However in the analysis
of the discrete symmetries and reality conditions in sect.~\ref{YM}
it will be important to use the solution $S^{sym}_{0\ga}$,
which is invariant under the corresponding discrete transformations.
In fact, the derivation of its explicit form
(\ref{S sym}) is one of the important results of this paper.
We expect that it is this solution
that is appropriate for the analysis of a locality in the model
as discussed in the subsequent sections
\footnote{When some restrictions on the class of transformations
are imposed by locality, the statement that different solutions
for $S_0$ belong to the same gauge equivalence class
should be reconsidered.}.

Now, let us turn to the equation (\ref{WS_0}). Since $dS_{0\ga}=0$, we get
\be
\label{Wt}
   [W_0, S_{0\ga}]_*=0 \,.
\ee
Thus, $W_0$ belongs to the subalgebra $A_S\subset A$ spanned by elements
that commute with $S_{0\ga}$, i.e. $A_S$ is the centralizer of $S_{0\ga}$.
For the case of $\nu = 0$, from (\ref{z,f}) it follows
that $A_S$ is the subalgebra of functions independent of $z$.
To find  $A_S$ for general $\nu$ we construct the elementary
generating elements $\ty_\ga$ commuting with $S^\pm_{0\ga}$
(\ref{S_0}) or $S^{sym}_{0\ga}$ (\ref{S sym}). The final result is
\be
\label{tilde y}
   \ty^\pm_\ga (z,y) = y_\ga+\nu (z_\ga \pm y_\ga)\int_0^1dt(t-1)
   e^{it(zy)}k \,,
\ee
\bee
\label{y sym}
   \ty^{sym}_\ga (z,y) & \!=\! & y_\ga + k \; {\nu\over 8} \;
     \int^1_{-1} ds (1-s)
     \exp\left\{\frac{i}{2}(s+1)(zy)\right\}  \, \left[ (y_\ga + z_\ga) \;
     \Phi \left({1\over 2}\,, \; 2\,; \; - k \; \mbox{ln} \,|s|^{\nu} \right)
        \right.  \nonumber \\
     & & \left. {} - (y_\ga - z_\ga) \;
     \Phi \left({1\over 2}\,, \; 2\,; \;  k \; \mbox{ln} \,|s|^{\nu}\right)
        \right] \,.
\eee
Remarkably, $\ty_\ga$ again obey the commutation relations of
the form (\ref{y mod}),
\be
\label{y com}
   [\ty_\ga,\ty_\gb]_*=
      2i\gep_{\ga\gb}(1+\nu k)\,,\quad
      \ty_\ga k=-k \ty_\ga \,.
\ee
It is easy to check that $[\ty^\pm_\ga\,,S^\pm_{0\gb}]_* =0$. The
fact that $[\ty^{sym}_\ga\,,S^{sym}_{0\gb}]_* =0$ is less trivial
(see Appendix B).
Note that as a by-product we have found that the deformed oscillator
algebra can be realized in terms of the embedding into the tensor
product of two Heisenberg-Weyl algebras equipped with
the generating element $k$.

Since $k$, $\psi_1$, and $\psi_2$ commute with $S_{0\ga}$,
the subalgebra $A_S$ is spanned by
the power series of $\ty_\ga$, $\psi_1$, $\psi_2$, and $k$,
i.e. its generic element has the form
\be
\label{A}
    A_S(z,y;\psi_{1,2},k)= \sum_{B,C,D=0}^1\sum_{n=0}^\infty \frac1{n!}\;
    A^{BCD}_{S\; \ga_1\ldots\ga_n}\; k^B \psi_1^C\psi_2^D \;
    \ty^{\ga_1}*\ldots*\ty^{\ga_n}\,,
\ee
where $A_{S\;\ga_1\ldots\ga_n}$ are totally symmetric multispinors
(i.e. we choose the Weyl ordering).
Note that due to the presence of $\psi_1$ and
$\psi_2$, $A_S$ is isomorphic to $Aq(2,\nu) \otimes Mat_2({\bf C})$.
According to (\ref{Wt}), $W_0$ should have a form (\ref{A}).

Since the commutation relations (\ref{y com}) have a form of the
deformed oscillator algebra (\ref{y mod}), one can use the
properties (\ref{T})-(\ref{oscom}) to build a solution
of the vacuum equations (\ref{WW_0}).
Namely, we can choose $W_0$ in the form analogous to (\ref{W gr}),
\be
\label{W_0}
    W_0=\go_0+\gl h_0\psi_1\,,\quad
       \go_0=\frac1{8i}\,\go_0^{\ga\gb}
                \{\ty_\ga, \ty_\gb\}_* \,,\quad
       h_0=\frac1{8i}\, h_0^{\ga\gb}
                 \{\ty_\ga, \ty_\gb\}_*\,.
\ee
Then (\ref{WW_0}) leads to the zero-curvature
and zero-torsion conditions (\ref{d omega}), (\ref{dh}) for
$\go_0^{\ga\gb}$ and $h_0^{\ga\gb}$, thus describing
AdS background geometry, so that the fields $\go_0^{\ga\gb}$ and
$h_0^{\ga\gb}$ can be interpreted, respectively, as the
background AdS Lorentz connection and dreibein of sect.~\ref{Prel}.
This completes construction of the background solution.

Let us emphasize that the form of the constraint (\ref{SS}) leads
in a rather non-trivial way
to the AdS background geometry via realization of the vacuum
centralizer $A_S$ in terms of the deformed oscillators $\ty_\ga$.

\section{Linearization}\label{LinEq}

Now we study the system (\ref{WW})-(\ref{SB})
perturbatively expanding it near the vacuum solution
(\ref{B_0}), (\ref{S sym}), and (\ref{W_0}) as
\be
\label{pert}
   B=B_0+B_1+\ldots \,,\qquad
   S_\ga=S_{0\ga}+S_{1\ga}+\ldots \,,\qquad
   W=W_0+W_1+\ldots \,.
\ee
This gives in the first order
\be
\label{WW_1}
   D_0\,W_1=0 \,,
\ee
\be
\label{WB_1}
   D_0\,C=0\,,
\ee
\be
\label{WS_1}
   D_0\, S_{1\ga}=[W_1, S_{0\ga}]_* \,,
\ee
\be
\label{SS_1}
   [S_{0\ga}, S_1{}^{\ga}]_*=-2i\, C*K \,,
\ee
\be
\label{SB_1}
   [S_{0\ga}\,,\,C]_*=0 \,,
\ee
where we denote $C=B_1$, and $D_0$ is the background covariant
adjoint differential, $D_0\,P=dP-W_0\wedge P+(-)^r P\wedge W_0$
for a $r$-form $P$.

{}From  (\ref{SB_1}), we see that $C$ takes its values in
$A_S$, i.e.
$C=C(\ty\,;\,k\,,\,\psi_{1,2})$. Inserting
\be
\label{decomp}
   C=C^{aux}(\psi_1)+C^{dyn}(\psi_1)\psi_2
\ee
into (\ref{WB_1}), one arrives at
\be
\label{C aux}
    D_0^L C^{aux}=\psi_1\gl[h_0,C^{aux}]_* \,,
\ee
\be
\label{C dyn}
    D_0^L C^{dyn}=\psi_1\gl\{h_0,C^{dyn}\}_*  \,,
\ee
where $D_0^L$ is the Lorentz covariant differential,
\be
   D_0^L C=dC-[\go_0,C]_* \,.
\ee
The equations (\ref{C aux}) and (\ref{C dyn}) coincide with
(\ref{aux}) and (\ref{dyn}), respectively.
Because the commutation relations for
the generating elements $\ty_\ga$ have the form (\ref{y com}),
from the results of sect.~\ref{OpReal} it follows
that the equations (\ref{WB_1}) and (\ref{SB_1})
lead in the sector of $C^{dyn}$ to the Klein-Gordon and Dirac
equations for spin-0 and spin-$\frac12$ fields with the masses (\ref{M}),
(\ref{M f}). As before, the rank-0 and 1 components of the expansion
(\ref{A}) for $C^{dyn}$ are identified with physical scalar and spinor
matter fields respectively.

As mentioned in sect.~\ref{OpReal}, the equations (\ref{C aux})
for the fields $C^{aux}$ describe an infinite set of auxiliary
(topological) fields,
each carrying no dynamical degrees of freedom (locally).

As a next step, we determine $S_{1\ga}$ from (\ref{SS_1}).
It is not difficult to integrate (\ref{SS_1}) perturbatively.
We however will apply a useful trick exploiting
the possibility to vary the free parameter $\nu$.
(All generating functions acquire
a non-trivial dependence on $\nu$ via the dependence on
the deformed oscillators $\tilde{y}$ (\ref{tilde y}).)
The advantage of this approach is that it works
for any vacuum solution $S_{0\ga}$, while the resulting concrete
formulae look very differently for different solutions like
$S^\pm_{0\ga}$ (\ref{S_0}) and $S^{sym}_{0\ga}$ (\ref{S sym})
and sometimes are not too illuminating.

With the aid of (\ref{SS_0}) and (\ref{SB_1}) one can easily check that
\be
\label{S_1}
   S_{1\ga}=C*\frac{\partial S_{0\ga}}{\partial\nu}+[S_{0\ga}, \chi]_*
\ee
is a solution of (\ref{SS_1}) with an arbitrary function
$\chi(z,y;\psi_{1,2},k)$. This ambiguity is a manifestation of the gauge
freedom (\ref{delta S}) at the linearized level with $\gvep_1 = - \chi$.
We therefore gauge fix $\chi=0$ by setting
\be
\label{S_1 0}
   S_{1\ga}=C*\frac{\partial S_{0\ga}}{\partial\nu}  \,.
\ee
By substituting this solution into (\ref{WS_1}) and making use
of (\ref{Wt}), (\ref{WB_1}), and (\ref{SB_1}), we get
\be
\label{W1t}
   \mbox{$ [C*\frac{\partial}{\partial\nu} W_0\,,\, S_{0\ga}]_*=
      [W_1\,,\,S_{0\ga}]_*  \,, $}
\ee
and, therefore,
\be
\label{W_1}
   W_1=\go(\ty; \psi,k)+C*\frac{\partial W_0}{\partial\nu}\,,
\ee
where an arbitrary one-form $\go(\ty; \psi_{1,2},k)$
takes values in $A_S$. It is this one-form $\go(\ty; \psi_{1,2},k)$
that is identified with the generating function for the HS gauge fields.
The substitution of (\ref{W_1}) into (\ref{WW_1}) gives
\be
\label{D-D}
   D_0\,\go = -D_0\,\left(C*\frac{\partial W_0}{\partial\nu} \right)\,.
\ee
The evaluation of the r.h.s. of this formula is analogous to the
derivation of (\ref{W1t}) and leads to
\be
\label{D W(C)}
   D_0\,\left(C*\frac{\partial W_0}{\partial\nu} \right)=
     C*\frac{\partial}{\partial\nu} (dW_0-W_0*\wedge W_0)=0 \,,
\ee
according to (\ref{WW_0}). Thus we conclude that
\be
\label{R=0}
   D_0\,\go = d\go-W_0 *\wedge \go-\go *\wedge W_0=0 \,,
\ee
i.e. there is no contribution
to the r.h.s. of the HS strengths in the first order
in $C$.

The first order corrections to the HS strengths
have been studied for the case of
massless matter fields with $\nu=0$ in \cite{Eq}, where
it was shown that such corrections are trivial
(can be compensated by some field redefinitions).
We generalize this conclusion to the case of an arbitrary $\nu$.
This is of course a desired result because the right hand sides of
the Chern-Simons dynamical equations for HS gauge fields can acquire
corrections due to some currents constructed from
the matter fields, but there is
no reason to expect non-trivial currents linear in physical fields.
We therefore conclude that, as expected, in the first order the system
(\ref{WW})-(\ref{SB}) describes free matter fields in some background
of the vacuum gauge fields (AdS background for the simplest
vacuum (\ref{W_0})).
This completes the analysis of the linearized equations. Two comments
are now in order.

The first is that at any order the perturbative analysis
of the system will reduce to some ${\partial\over \partial x^\mu}$ and
$[S_0 \,,\ldots ]_*$ equations of the type
(\ref{WW_1})-(\ref{SB_1}) with the r.h.s.-s determined via solutions
in the lowest orders. This system is overdetermined but  consistent
due to the general consistency properties of the full system with
respect to farther ${\partial\over \partial x^\mu}$ differentiations and
taking the commutators $[S_\ga, \ldots]_*$. Some specificities (but no
inconsistencies) can only take place for the degenerate points
$\nu=2l+1$, $l\in {\bf Z}$ discussed in sect.~\ref{OpReal} and Appendix B.

The second comment is that the r.h.s.-s are always well defined.
The point is that our formulae contain some non-polynomial expressions
in the variables $y_\ga $ and $z_\gb$ (cf. (\ref{S_0}), (\ref{tilde y})),
and it is not guaranteed {\it a priori}
that $*$-products of such expressions do not contain divergencies.
The fact that all perturbative
computations are well defined is based on the theorem proved in \cite{Prop},
which guarantees the regularity of all our expressions provided that the
original variables belong to a certain class of regular functions.
In Appendix A, we reformulate this theorem
in the form appropriate to the analysis in this paper and show
that $S_{0\ga}$, $\ty_\ga$ and any functions of $\ty_\ga$ of the type
(\ref{A}) belong to the appropriate regularity class.

\section{Non-Local Integrating Mapping}\label{Integr}

A remarkable property of the non-linear system
(\ref{WW})-(\ref{SB}) is that it admits a flow that allows one
to express constructively its solutions in terms of
those of the linearized problem. This construction is a generalization of
the field redefinitions (\ref{S_1 0}), (\ref{W_1}) used in sect.~\ref{LinEq}
to show that HS gauge field strengths do not have
nontrivial sources linear in fields.

We introduce a parameter $\tau$  as
an additional evolution parameter. The meaning of $\tau$ is that
it serves as an expansion parameter in powers of interactions,
i.e. as an effective coupling constant. The idea is that
the evolution flow in $\tau$ will allow us to express the dynamical
fields of the non-linear problem with $\tau \neq 0$ via those of
the free problem with $\tau =0$. Let us note that this method has some
similarity with the method of evolution over a coupling constant developed
in \cite{kir} in a different context.

We therefore assume that $W=W(\tau)$, $B=B(\tau)$ and
$S_\ga=S_\ga(\tau)$ and introduce a shifted variable
${\cal B}(\tau)$,
\be
\label{fr}
   B(\tau)=\nu+\tau {\cal B}(\tau) \,,
\ee
where $\nu$ is some constant parameter, the vacuum value of the
field $B$, while ${\cal B}$ is assumed to be a fluctuational part.
The system (\ref{WW})-(\ref{SB}) acquires the form
\be
\label{WWe}
   dW=W*\wedge W  \,,
\ee
\be
\label{WBe}
   d{\cal B}=W*{\cal B}-{\cal B}*W \,,
\ee
\be
\label{WSe}
   dS_\ga=W*S_\ga-S_\ga*W          \,,
\ee
\be
\label{SSe}
   S_\ga*S^\ga=-2i(1+\nu K+\tau {\cal B}*K)  \,,
\ee
\be
\label{SBe}
   S_\ga*{\cal B}={\cal B}*S_\ga \,.
\ee
Note that the parameter $\tau$
falls out of all the equations except for (\ref{SSe}).

Now one observes that for the limiting case $\tau=0$, the system
(\ref{WWe})-(\ref{SBe}) reduces to the free one. Indeed, setting
$$
   \nu=B_0\,,\quad {\cal B}(\tau=0)=B_1\equiv
       C\,,\quad W(\tau=0)=W_0\equiv \go\,, \quad S_\ga(\tau=0)=S_{0\ga} \,,
$$
we see that at $\tau=0$ the system (\ref{WWe})-(\ref{SBe}) has the
form of the vacuum system
(\ref{B_0})-(\ref{SS_0}) plus the linear equations
(\ref{WB_1}), (\ref{SB_1}) for the matter fields $C$.
Hence, after $S_{0\ga}$ is excluded, we arrive at the free system
\be
\label{of}
    d\go = \go * \wedge \go
\ee
\be
\label{Cf}
    dC = \go * C - C * \go \,,
\ee
where $\go=\go(\ty)$ and $C=C(\ty)$ belong to $A_S$.
Let us note that the situation here is similar to that
with contractions of Lie algebras. For all values of $\tau \neq 0$,
the systems of equations (\ref{WWe})-(\ref{SBe}) are pairwise equivalent
since the field redefinition (\ref{fr}) is non-degenerate.
On the other hand, although the field redefinition (\ref{fr}) degenerates
at $\tau=0$, the system of equations (\ref{WWe})-(\ref{SBe}) still
makes sense, describing the free field dynamics.

Let us now define a flow with respect to $\tau$,
\be
\label{ps 1}
   \frac{\partial W}{\partial\tau}=
     (1-\mu)\;{\cal B}*\frac{\partial W}{\partial\nu}
     +\mu\; \frac{\partial W}{\partial\nu}*{\cal B} \,,
\ee
\be
\label{ps 2}
  \frac{\partial {\cal B}}{\partial\tau}=
   (1-\mu)\;{\cal B}*\frac{\partial {\cal B}}{\partial\nu}
   +\mu\; \frac{\partial {\cal B}}{\partial\nu}*{\cal B}   \,,
\ee
\be
\label{ps 3}
   \frac{\partial S_\ga}{\partial\tau}=
    (1-\mu)\; {\cal B}*\frac{\partial S_\ga}{\partial\nu}
    +\mu\; \frac{\partial S_\ga}{\partial\nu} * {\cal B} \,,
\ee
where $ \mu $ is an arbitrary parameter.
By applying $\frac{\partial}{\partial\tau}$ to the both sides
of eqs.(\ref{WWe})-(\ref{SBe}) one concludes that for any $ \mu $
the system (\ref{ps 1})-(\ref{ps 3}) is compatible with
(\ref{WWe})-(\ref{SBe}). Therefore, solving the system
(\ref{ps 1})-(\ref{ps 3}) with the initial data
\be
\label{it}
   {\cal B}(\tau=0)=C\,,\quad W(\tau=0)=\go\,,
   \quad S_\ga(\tau=0)=S_{0\ga} \,,
\ee
we can express solutions of the full nonlinear system at $\tau=1$
via solutions of the free system at $\tau =0$. This approach is
very efficient at least perturbatively and allows one to derive
order by order the relevant field redefinitions.
In particular, it is easy to see that at $\mu = 0$ it leads directly to
the solution (\ref{S_1 0}) used in sect.~\ref{LinEq}
to show that the HS gauge field strengths do not admit
non-trivial sources linear in fields.
Note that even at the linearized level it is a
complicated technical problem to find a form of  an
appropriate solution
without using the flow (\ref{ps 1})-(\ref{ps 3}).

The flows (\ref{ps 1})-(\ref{ps 3}) at different $\mu$ develop within
the same gauge equivalence class.
To see that any variation of $\mu$ is induced
by some gauge transformation one should
find a gauge parameter $\gvep$ such that
\be
\label{mu ev}
   {\partial W\over \partial\mu}  = D\gvep \,,\qquad
   {\partial {\cal B}\over \partial\mu}  = [\gvep\,,\,{\cal B}]_* \,,\qquad
   {\partial S_\ga\over \partial\mu}  = [\gvep\,,\,S_\ga]_* \,,\qquad
\gvep (\tau =0) =0\,,
\ee
where $D\gvep = d\gvep - [W\,,\,\gvep]_*$. The compatibility
condition of (\ref{mu ev})
 with (\ref{ps 1})-(\ref{ps 3}) is satisfied if
\be
\label{eq e}
   \frac{\partial \gvep}{\partial\tau} =
   \frac{\partial {\cal B}}{\partial\nu} +
   (1-\mu)\; {\cal B} * \frac{\partial \gvep}{\partial\nu}
   +\mu\; \frac{\partial\gvep}{\partial\nu} * {\cal B}  \,,
\ee
which condition just fixes a $\tau$-dependence of $\gvep$.
Thus, one is free to choose any value of $\mu$. There are three
most interesting cases: $\mu =0$ (left), $\mu =1$ (right) and
$\mu = \frac{1}{2}$ (symmetric).

Important comment is that one has to be careful in making statements
on the locality of the mapping induced by the flow
(\ref{ps 1})-(\ref{ps 3}). Indeed, although it does not contain
explicitly space-time derivatives, it contains them implicitly
via highest components $C_{\ga(n)}$ of the generating function
$C(\tilde{y})$, which are identified with the highest derivatives
of the matter fields by the equations (\ref{Cf})
(see the chains of equations (\ref{chainbos+-}), (\ref{chainferm+-})
which are the component forms of (\ref{Cf})).
For example, the equation (\ref{ps 2}) at $\mu=0$ in the zero order
in $\tau$ reads
\be
\label{ps 22}
   \frac{\partial}{\partial\tau}{\cal B}_1(z,y)=
        C(\ty)*\frac{\partial C(\ty)}{\partial\nu} \,.
\ee
Because of using the $*$-product, for each fixed rank multispinorial
component of the l.h.s. of this formula there appears, in general,
an infinite series involving bilinear combinations of the components
$C_{\ga(n)}$ with all $n$ on the r.h.s. of (\ref{ps 22}).
Therefore, the r.h.s. of (\ref{ps 22}) effectively involves
all order space-time derivatives, i.e. the transformation laws
(\ref{ps 1})-(\ref{ps 3}) can effectively describe some non-local
transformations. This means that we cannot consider
the system (\ref{WW})-(\ref{SB}) as locally equivalent to the free
system (\ref{of}), (\ref{Cf}). Instead we can only claim that there
exists a non-local mapping between the free and nonlinear system.
This mapping is reminiscent of the Nicolai mapping in supersymmetric
models \cite{Nic} and the B\"acklund mapping in the integrable systems.

At the linearized level, however, the transformations
induced by the integrating flow (\ref{ps 1})-(\ref{ps 3}) turn out to be
local for the following simple reason. In this case, all field
redefinitions are linear in the matter fields $C$. They can
contain HS gauge fields $\go$, but in the linear
approximation only the vacuum (zero-order) part of the gauge fields can
contribute. This part is nothing else as the background gravitational
1-forms (\ref{W_0}) which are bilinear in the auxiliary variables
$\tilde{y}_\ga$. As a result, the transformations for
physical fields, induced by (\ref{ps 1}), are always polynomial
in $\tilde{y}_\ga$. So, they can contain at most
a finite number of the coefficients $C_{\ga(n)}$.
This is equivalent to the statement that
the linearized field transformations contain only a finite number of
space-time derivatives and therefore are local.
Thus, the statement that the equations of motion for HS
fields do not acquire sources linear in the matter fields is the well
defined local statement. This is not expected to be the case for
the second-order analysis which should cover the case of bilinear
HS currents constructed from the matter fields. To illustrate
this point we consider in Appendix C an example of $d3$ gravity
interacting with some matter source.

\section{Extended Symmetries and Truncations}\label{YM}

\subsection{Inner Symmetries}

The HS dynamical systems admit a
natural extension to the case with non-Abelian
internal (Yang-Mills) symmetries, as was first discovered
in \cite{Ann, KV1} for the $d4$ case
\footnote{Recently, a particular case of this extension \cite{OP1}
was used in \cite{sez} to describe $d4$ $N=8$ HS supergravity.}.
The key observation is that the system (\ref{WW})-(\ref{SB})
remains consistent if components of all fields take their values
in an arbitrary associative algebra $M$ with a unit element $I_M$,
i.e. the fields $W$, $B$, and $S_\ga$ take their values
in the algebra ${\cal A}^{ext} = {\cal A}\otimes M$, where ${\cal A}$
is the associative algebra with the general element (\ref{gexp}).
The extended HS algebra $A^{ext}$ is then identified with
the $\rho$-independent part of ${\cal A}^{ext}$, while the fields $S_\ga$
are assumed as before to be linear in $\rho$.
The algebra ${\cal A}$ admits a natural embedding
${\cal A}\to {{\cal A} \otimes I_M}$.
The gravitational sector is associated with
$A\sim A\otimes I_M$ and therefore commutes with $M\sim I_A\otimes M$,
where $I_A$ is the unit element of $A$. Therefore, $M$ describes
internal symmetries in the model. So far, we have considered
 the simplest Abelian case. Now we turn to
a general situation with finite-dimensional inner symmetries.
For the case of semisimple finite-dimensional inner symmetries,
the only possibility is to identify $M$ with some matrix algebra.
It is convenient to start with the complex algebra $M=Mat_n ({\bf C})$,
imposing the reality conditions afterwards.

Thus we set
$$
   W(z,y;\psi_{1,2},k)\to W_i{}^j(z,y;\psi_{1,2},k) \,,\quad
   B(z,y;\psi_{1,2},k)\to B_i{}^j(z,y;\psi_{1,2},k) \,,
$$
$$
   S_\ga(z,y;\psi_{1,2},k,\rho)\to
   S_{\ga,\,i}{}^j(z,y;\psi_{1,2},k,\rho)  \,,
$$
where the matrix indices $i,j = 1\ldots n$ do not affect
Grassmann properties (\ref{pWB}), (\ref{pS}).

The vacuum solutions (\ref{B_0}), (\ref{S_0}), (\ref{S sym}), (\ref{W_0})
give rise to the vacuum solutions of the extended system
via embedding ${\cal A}\to {{\cal A}\otimes I_M}$, i.e.
\be
\label{}
  W_{0,\;i}{}^j = W_0 \;\gd_i{}^j\,,\qquad
  B_{0,\;i}{}^j = B_0 \;\gd_i{}^j\,,\qquad
  S_{0\ga,\;i}{}^j = S_{0\ga} \;\gd_i{}^j \,.
\ee
The infinite-dimensional symmetry algebra $l^{ext}$ of the model is
the Lie superalgebra defined via supercommutators
in $A^{ext} = A\otimes Mat_n({\bf C})$. More precisely, we need
a real form of the original complex Lie superalgebra $l^{ext}$
that is compatible with the unitarity of the model.
Let us denote this real superalgebra as $hu(n|4)$, where $n$
indicates the factor of $Mat_n({\bf C})$ in the original algebra $A^{ext}$,
while 4 is for the four generating elements $z_\ga, y_\ga$.

The appropriate reality conditions can be imposed with the aid of
the involution defined as the mapping (\ref{inv}), when acting on ${\cal A}$,
and as an ordinary Hermitian conjugation on $Mat_n ({\bf C})$,
\be
   (a_i{}^j )^\dagger = \bar{a}_j{}^i \,.
\ee
This uniquely defines some
involution $\dagger$ of the full algebra ${\cal A}^{ext}$.
Then, the real HS Lie superalgebra $hu(n|4)$ is extracted by imposing
the condition\footnote{More details on the relationship between
involutions, conjugations and real forms of complex (super)algebras can
be found in \cite{OP1}.}
\be
\label{hermit alg}
    a^{\dagger}=-i^{\pi(a)}\,a
\ee
on $a\in l^{ext}$, where $\pi =0$ or $1$ is the boson-fermion
parity given by (\ref{pWB}).

The appropriate (anti)hermiticity conditions for the variables
$W\,,\,B\,,\,S_\ga$ consistent with the system (\ref{WW})-(\ref{SB}) read
\be
\label{hermit}
   W^{\dagger} = -W\,,\quad S_\ga^{\dagger} = -S_\ga \,,\quad
              B^{\dagger} = B \,,
\ee
where an additional sign factor for $B$ is due to the factor of $i$
on the r.h.s. of (\ref{SS}).
In these formulae, it is assumed that $\dagger$ is also the involution
of the Grassmann algebra in which the components of the fields
$W\,,\,B\,,\,S_\ga$ take their values, i.e. it reverses an order of
Grassmann factors.

Note that this definition of the reality conditions is compatible
with the vacuum solution $B_0$, $W_0 (\ty^{sym}_\ga)$ and
$S^{sym}_{0\ga}$ (cf. (\ref{S sym}), (\ref{y sym})),
but not with the solutions with
$S^+_{0\ga}$ and $S^-_{0\ga}$ which are conjugated to each other.
Therefore, from now on we choose $S_{0\ga} = S^{sym}_{0\ga}$.

\subsection{Truncations by Automorphisms}\label{Auto}

Now we study consistent truncations of the extended system
(\ref{WW})-(\ref{SB}) induced by its various involutive symmetries.
Let $\eta$ be some involutive symmetry
of the system (\ref{WW})-(\ref{SB}), i.e. $\eta^2 = I$.
Then the truncation conditions are
\footnote{The involutivity condition
is not essential. This is just a particular case of gauging some
discrete symmetry group $\Gamma$. In our case $\Gamma$ will be a product
of some $Z_2$ factors.}
\be
\label{trunc}
    \eta(A_i) = A_i \,,\quad A_i = (W_\mu\,,\,B\,,\,S_\ga) \,.
\ee
Such a truncation is consistent in the standard
sense that every solution of the truncated system is a solution of the
full system. Indeed, it is equivalent to setting to zero the variables
$A_i^{odd}$ from the odd subspace, $\eta(A_i^{odd}) = - A_i^{odd}$.

In this section, we consider the truncations of the system
(\ref{WW})-(\ref{SB}) induced by automorphisms of the associative
algebra ${\cal A}^{ext}$.
The following automorphisms of ${\cal A}^{ext}$ will be important,
\be
\label{ap}
    f[a(z,y;\rho,k,\psi_{1,2})] = a(-z,-y;-\rho,k,\psi_{1,2}) \,,
\ee
\be
\label{psi inv}
    \phi [a(z,y;k,\psi_1 ,\psi_2 )] = a(z,y;k,\psi_1 ,-\psi_2 ) \,,
\ee
and the similarity transformation $\kappa$ acting on
$(n+m)\times (n+m)$ matrices $M$ as follows,
\be
\label{kappa}
    \kappa(M)=UMU^{-1} \,,\quad
    \begin{array}{rcl}
          U & = & \left( \begin{array}{ccc}
                                I_n   &    0 \\
                                0    &    -I_m
                         \end{array}
                  \right)  \,.
    \end{array}
\ee

Following to \cite{KV1} we will use the algebra
$hu(n+m|4)$ as a starting point for the derivation of a series of
algebras realized in various HS models that can be obtained by virtue of
truncations of the equations (\ref{WW})-(\ref{SB}).
The mappings (\ref{ap}), (\ref{psi inv}) and (\ref{kappa}) are mutually
commuting involutive automorphisms of $A^{ext}$ and, as a consequence,
of $hu(n+m|4)$. One can consider their compositions $\phi^\ga f^\gb \kappa$
with $\ga, \gb = 0$ or $1$, which are involutive automorphisms as well
\footnote{The truncations based on the
automorphisms of the form
$\phi^\ga f^\gb $ are not interesting since they eliminate
either all fermions or dynamical matter bosons,
or all the dynamical matter degrees of freedom.}.
These automorphisms induce involutive symmetries of the full system
(\ref{WW})-(\ref{SB}). The corresponding truncation conditions are
\be
\label{pkW}
    \phi^\ga f^\gb \kappa \; (W) = W \,,\qquad
    \phi^\ga f^\gb \kappa \; (S_\gga) = S_\gga \,, \qquad
    \phi^\ga f^\gb \kappa \; (B) = B \,.
\ee

Let us stress that we confine ourselves to only the truncations
that do not affect the vacuum solutions (\ref{B_0}), (\ref{S sym}),
(\ref{W_0}) since otherwise the resulting truncated
equations will not admit a perturbative interpretation
(this is why we do not consider the automorphism
similar to (\ref{psi inv}) that changes a sign of $\psi_1$).

Now let us analyze the gauge symmetry algebras of the truncated systems
that are spanned by the elements $a$ satisfying the conditions
\be
\label{trcon}
    \phi^\ga f^\gb \kappa \; (a) = a \,,\qquad a \in hu(n+m|4) \,.
\ee
First of all one observes that the truncated algebra acquires
an involutive central element $\psi_1^\ga k^\gb U$ which
commutes to every element satisfying (\ref{trcon}).
This means that the truncated algebra decomposes
into a direct sum of two subalgebras projected out by the projectors
\be
\label{proj}
    {\cal P}^\pm = {(1\pm \psi_1^\ga k^\gb U)\over 2} \,,\qquad
    {\cal P}^\pm l^\pm = l^\pm \,.
\ee
Let the corresponding subalgebras $l^\pm$ be called
$hu^\pm_{\ga\gb}(n,m|4)$.
The full system of equations also decomposes into two independent
(mutually non-interacting) subsystems singled out by the conditions
${\cal P}^\pm (A^\pm_i )= A^\pm_i$,
$ A^\pm_i = (W^\pm_\mu\,,\,B^\pm\,,\,S^\pm_\ga).$
We will refer to these subsystems as based on
$hu^\pm_{\ga\gb}(n,m|4)$.

The case $\ga = \gb =0 $ is trivial with
$hu^+_{00}(n,m|4) \sim hu(n|4)$ and $hu^-_{00}(n,m|4) \sim hu(m|4)$.

The case $\ga =0$, $\gb =1$ is similar to the $d4$ algebras
considered in \cite{KV1}. The
algebras $hu^+_{01}(n,m|4) \sim hu^-_{01}(m,n|4)$
are spanned by the matrices from  $hu(n+m|4)$
with bosons in the diagonal blocks $n\times n$ and $m\times m$,
and fermions in the off-diagonal blocks $n\times m$ and $m\times n$.
The Klein operator $k$ should be replaced by the matrix $U$, according
to the projection conditions (\ref{proj}).
As a result we arrive at the two-parametric set of consistent models
with HS gauge symmetry algebras that are
the subalgebras of $hu(n+m|4)$ extracted by
the condition (\ref{trcon}) and the projectors ${(1\pm kU)\over 2}$.
In notations of \cite{KV1} these algebras are $hu(2n,2m|4)$
\footnote{The doubling of $n$ and $m$ is due to the presence of the
elements $\psi_{1,2}$ which generate the Clifford algebra
isomorphic to the  $2\times 2$ matrix algebra.}.

The case $\ga = 1$, $\gb = 0$ is analogous to the massless
case considered in \cite{Eq} (where the Klein operator did not appear).
The corresponding algebras $hu^+_{10}(n,m|4) \sim hu^-_{10}(m,n|4)$
are spanned by the matrices
from $hu(n+m|4)$ with $\psi_2$-independent diagonal blocks
$n\times n$ and $m\times m$, and linear in $\psi_2$
off-diagonal blocks $n\times m$ and $m\times n$.
For these algebras, $\psi_1$ should be replaced by the matrix $U$
according to the form of the projector (\ref{proj}).
The field C (\ref{decomp}) acquires the form
$
    \left( \begin{array}{rcl}
                   C^{aux}   &   C^{dyn}   \\
                   C^{dyn}   &   C^{aux}
            \end{array}
    \right)
$,
while the background fields become $W_0=\go_0 I + \gl h_0 U$,
where $I$ is the $(n+m)\times (n+m)$ unit matrix.

The mixed case $\ga =1$, $\gb =1$ can be described analogously.
The elements of the algebras
$hu^+_{11}(n,m|4) \sim hu^-_{11}(m,n|4)$ are the matrices
from $hu(n+m|4)$ with $\psi_2$-independent bosons and linear in
$\psi_2$ fermions in the diagonal blocks $n\times n$ and $m\times m$, and
with $\psi_2$-independent fermions and linear in $\psi_2$ bosons
in the off-diagonal blocks $n\times m$ and $m\times n$.
In this case, $\psi_1$ should be replaced by $kU$,
in accordance with the projection condition (\ref{proj}).

In all three non-trivial cases with $\ga + \gb \neq 0$, the isomorphisms
$hu^\pm_{\ga\gb}(n,m|4) \sim hu^\pm_{\ga\gb}(m,n|4)
\sim hu^\mp_{\ga\gb}(m,n|4)$ take place.
For this reason, in what follows the signs ``$\pm$'' will be omitted.
In general, algebras with different pairs $n, m$ and $\ga, \gb$
($\ga + \gb\neq 0$) are pairwise non-isomorphic and lead to the different
HS models. At $n=m$, all these algebras are isomorphic to the
original algebra, $hu_{\ga\gb}(n,n|4) \sim hu(n|4)$, $\ga, \gb = 0$ or 1
\footnote{One can make sure of this taking into account that the doublings
of elements of the algebras due to the operators $k$ and $\psi_1$
turn out to be equivalent to the doubling of the matrix blocks for even
and odd elements separately, with appropriate replacements $k\to U$
or $\psi_1\to U$, or $k\psi_1\to U$.}.

The Yang-Mills subalgebras of the infinite-dimensional algebras
described above are spanned by the elements
independent of $z_\ga$, $y_\ga$ and $\psi_2$
(to ensure commutativity with the space-time symmetry algebra
$o(2,2)$ (\ref{W_0})). For all three non-trivial cases with
$\ga + \gb \neq 0$, the resulting Yang-Mills subalgebra is
$u(n)\oplus u(n)\oplus u(m)\oplus u(m)$
(the doubling is due to the presence of either $\psi_1$
or $k$, ($(\psi_1)^2=k^2 =1$)).

\subsection{Truncation by Antiautomorphism}\label{Anti}

As shown in \cite{Ann, KV1} for the $d4$ case and in \cite{Eq} for
the $d3$ case, further truncations of the extended systems
that lead to the orthogonal and symplectic Yang-Mills algebras
can be constructed with the aid of antiautomorphisms of the
original associative algebra ${\cal A}^{ext}$.
Every antiautomorphism $\sigma$ of $A^{ext}\subset {\cal A}^{ext}$
(i.e. a mapping with the properties (\ref{pr1}), (\ref{pr2}))
induces some automorphism $\tau$ of the Lie superalgebra
$l^{ext}$ (and of $hu_{\ga\gb}(n,m|4)$),
according to the following general rule
\be
\label{tausi}
    \tau(a) = -i^{\pi(a)} \; \gs(a) \,
\ee
(for more details on the relationship between antiautomorphisms
and automorphisms of Lie superalgebras see e.g. \cite{OP1}).
A subalgebra of $hu_{\ga\gb}(n,m|4)$ can now be extracted
by the condition $\tau(a)=a$.

To define some antiautomorphism $\gs$ on
${\cal A}^{ext} = {\cal A}\otimes Mat_n({\bf C})$ it suffices to define it
separately on ${\cal A}$ and $Mat_n({\bf C})$.
We define $\gs$ on ${\cal A}$ according to (\ref{sigma}).
To define how $\gs$ acts on $Mat_{n+m}({\bf C})$ one can use the fact that
antiautomorphisms $\gs^h$ of $Mat_{n+m}({\bf C})$ are induced by non-degenerate
bilinear forms $h(x,y) = H^{ij} x_i y_j$ on some representation space
of $Mat_{n+m}({\bf C})$, $h(x,My) = h(\gs^h(M)x,y)$, i.e.
\be
\label{sM}
    \gs^h(M_i{}^j) = H^{jk} M_k{}^l H^{-1}_{li} \,,\qquad
    \forall M\in Mat_{n+m}({\bf C}) \,.
\ee
Let us assume that
\be
\label{specH}
    \begin{array}{rcl}
              H & = & \left( \begin{array}{ccc}
                                    H_n   &    0 \\
                                    0     &    H_m
                             \end{array}
                      \right)  \,,\qquad
    \end{array}
    H_n^{ij}=\pm H_n^{ji} \,,\qquad H_m^{ij}=\pm H_m^{ji} \,.
\ee
The block-diagonal form of $H$ will be important to guarantee
that the vacuum solution satisfies the truncation condition.
It is important to note that $n\times n$ and $m\times m$ blocks
$H_n$ and $H_m$ are required to have the same symmetry properties
\footnote{In agreement with the analysis of the $d4$ case in \cite{KV1},
relaxing this condition does not lead to new possibilities.
Indeed, at $\ga+\gb \neq 0$, the choice of $H_n$ and $H_m$ with different
types of symmetry leads to  $\tau^2=\kappa$ (\ref{kappa}).
This is equivalent to
 setting $\kappa=1$ in (\ref{pkW}) and gives rise to the degenerate systems
truncated by the automorphisms $\phi^\ga f^\gb$ discussed in the footnote
in subsect.~\ref{Auto}. }.

The automorphism $\tau$ gives rise to the symmetry $\eta$ of the full
system,
\be
\label{ruleW}
     \eta(W_\mu) = -i^{\pi (W)} \; \gs(W_\mu) \,,
\ee
\be
\label{ruleB}
     \eta(B) = i^{\pi (B)} \; \gs(B) \,,
\ee
\be
\label{ruleS}
     \eta(S_\ga) = i^{\pi (S) +1} \; \gs(S_\ga) \,,
\ee
where $\pi = 0$ or $1$ is given by (\ref{pWB}), (\ref{pS}).
It is assumed here that $\sigma$ does not affect the Grassmann field
coefficients (i.e. the order of the Grassmann factors is not reversed).
The truncation is then performed via imposing the condition
(\ref{trunc}). It is important that these conditions leave invariant
the vacuum solutions $B_0 \otimes I_M$ (\ref{B_0}),
$S^{sym}_{0\ga} \otimes I_M$ (\ref{S sym}) and
$W_0(\ty^{sym}_\ga) \otimes I_M$ (\ref{W_0}).

In the spin-1 sector $W^{YM}_\mu$
(i.e. a part of $W_\mu$ independent of $z,y$, and $\psi_2$),
the condition $\eta(W_\mu) = W_\mu$ reads $\gs^h(W^{YM}_\mu) = - W^{YM}_\mu$
and means that the fields $W^{YM}_\mu$ correspond to the subalgebra that
leaves the (skew)symmetric bilinear form $h$ invariant.
Thus, the truncations induced by the antiautomorphisms $\sigma$
reduce the unitary Yang-Mills subalgebras to either orthogonal
algebras $o(n)\oplus o(n)\oplus o(m)\oplus o(m)$ ($H$ is symmetric)
or symplectic algebras $usp(n)\oplus usp(n)\oplus usp(m)\oplus usp(m)$
($H$ is skewsymmetric and $n$ and $m$ are even).

Let us denote the respective subalgebras of $hu_{\ga \gb}(n,m | 4)$
as $ho_{\ga \gb}(n,m | 4)$ and \\ $husp_{\ga \gb}(n,m | 4)$,
in analogy with the notation of \cite{KV1}.
It is worth to mention that the role of the antiautomorphism $\gs$
is analogous to that of the twist (orientation)
operator in the superstring theory \cite{GSW}.

\section{Global (Super)Symmetries}\label{GlobSym}

\subsection{Global Symmetries}

The system (\ref{WW})-(\ref{SB}) is explicitly
invariant under the  HS gauge transformations
(\ref{delta W})-(\ref{delta S}).
Fixation of the vacuum solution (\ref{B_0}), (\ref{S sym}), (\ref{W_0})
breaks this local symmetry down to some global symmetry, the symmetry
of the vacuum. This global symmetry is also a symmetry of
the linearized system (\ref{WW_1})-(\ref{SB_1}) and, as a consequence,
of the equations (\ref{C aux}), (\ref{C dyn}). It is generated
by the parameter $\gvep_{gl}(x)$ obeying the conditions
\be
\label{e gl1}
    d\gvep_{gl}=[W_0\,,\,\gvep_{gl}]_*\,,
\ee
\be
\label{e gl2}
    [\gvep_{gl}\,,\,S_{0\ga}]_*=0  \,,
\ee
which follow from the requirements that $\gd W_0=0$ and $\gd S_{0\ga}=0$
($\gd B_0=0$ holds trivially), i.e. $\gvep_{gl}$ belongs to the stability
subalgebra of the vacuum solution
\footnote{Here it is important to note that, due to the zero-curvature
conditions (\ref{WW_0}), general coordinate transformations of the
vacuum solution reduce to some gauge transformations according to
the relations $\gd_\xi A_\nu = D_\nu \gvep + \xi^\mu R_{\mu\nu} $,
where $\gd_\xi$ is an infinitesimal diffeomorphism with a parameter $\xi^\mu$
and $\gvep=\xi^\mu A_\mu$.}.

The condition (\ref{e gl2}) implies that $\gvep_{gl}$ belongs to
 $A_S^{ext} = A_S \otimes Mat_n({\bf C})$,
i.e. $\gvep_{gl}$ is of the form (\ref{A}) with the matrix-valued
coefficients. The equation (\ref{e gl1}) fixes a dependence of $\gvep_{gl}$
on the space-time coordinates $x^\mu$ in terms of the initial data
$\gvep_{gl}(x_0)=\gvep_{gl}^0$ at any space-time point $x_0$
(in some neighborhood of $x_0$).

According to (\ref{A}), $A_S^{ext}$ contains elements
linear in $\psi_2$. {}From the definition (\ref{decomp}), it follows
that the symmetry parameters linear in $\psi_2$
mix the auxiliary fields $C^{aux}$ and the dynamical fields $C^{dyn}$.
As mentioned in sect.~\ref{OpReal}, $C^{aux}$
describes an infinite set of the auxiliary fields, each having at most
a finite number of degrees of freedom. This means that $C^{aux}$ decomposes
into a sum of finite-dimensional representations of
the space-time symmetry algebra $o(2,2)$. Since $o(2,2)$ is non-compact,
it has no non-trivial finite-dimensional unitary representation
and, therefore, these modes cannot survive in any unitary theory.
In terms of solutions of the field equations the point is that
the auxiliary fields do not admit non-trivial solutions after imposing
appropriate boundary conditions at space infinity.
Thus we conclude that the unitarity requirement rules out
the symmetry with the parameters $\gvep_{gl}$ linear in $\psi_2$.
So, similarly to the $d=4$ case \cite{FV2},
we will analyze the subalgebra $l^g$ of the full global
symmetry superalgebra that is spanned by the $\psi_2$-independent elements
\footnote{One can factor out a trivial  center spanned by constant
 parameters $\gvep_{gl}$.}
and admits unitary representations.

Since $l^g$ does not contain $\psi_2$, the element $\psi_1$
becomes central. As a result, $l^g$ decomposes into a direct sum of the two
subalgebras with the aid of the projectors $\frac{1}{2}(1\pm\psi_1)$,
i.e. $l^g = l^{g+} \oplus l^{g-}$. As $\psi_1$ is a coefficient
in front of the dreibein in (\ref{W_0}), for the linearized equations
to admit sensible flat limit one has to require that the two components
$l^{g+}$ and $l^{g-}$ be isomorphic to each other. Let us now
analyze what global symmetry algebras result from the gauge algebras
$hu_{\ga \gb}(n,m | 4)$, $ho_{\ga \gb}(n,m | 4)$ and
$husp_{\ga \gb}(n,m | 4)$
defined in the previous section.

In the case $\beta=0$ the two choices of $\ga$ lead to different results.
The case with $\ga=1$ is more general. {}From the projection
condition (\ref{proj}) one easily finds that the resulting global symmetry
algebra $hu_0 (n,m | 2) = hu_0(n,0 | 2) \oplus hu_0 (0,m | 2)$,
i.e. it has a form
$
   \begin{array}{rcl}
               \left( \begin{array}{rcl}
                                    A   &   0  \\
                                    0   &   B
                      \end{array}
                        \right)
    \end{array}
$
with $n\times n$ and $m\times m$ diagonal blocks $A$ and $B$ which
depend on the spinor variables $\tilde{y}$ and $k$. According
to (\ref{proj}), $\psi_1$ is identified here with the matrix $U$=
$
    \left( \begin{array}{rcl}
                   1   &   0   \\
                   0   &   -1
            \end{array}
    \right)
$. Therefore the dreibein is also proportional to $U$ and the flat limit
is possible only in the case with $n=m$ when left and right sectors
are symmetric. It is easy to see that if one starts with the case
$\ga = 0$ then the resulting global algebra is either
$hu_0 (n,n|2)$ or $hu_0 (m,m|2)$ depending on what sign
is chosen in (\ref{proj}). Note that this class of algebras
is analogous to that considered in \cite{Eq} for the $d=3$ massless
case.

In the case of $\gb = 1$ one arrives at the same result for the
both choices of $\ga$. Here $k$ is identified with $\pm U$ according to
(\ref{proj}), while $\psi_1$ should be kept as an independent generating
element. Therefore the resulting algebra $hu_1 (n,m|2)$ is left-right
symmetric and the corresponding linearized equations of motion
admit sensible flat limit for all $n$ and $m$. Here the bosonic symmetries
live in the diagonal blocks while the fermionic ones in the off-diagonal
blocks. These algebras are analogous to the $d=4$ global symmetry algebras
considered in \cite{KV1}.

The algebras
$ho_\beta (n,m|2)$ and $husp_\beta (n,m|2)$ are defined analogously.

\subsection{$N=2$ SUSY}\label{N=2}

In the case of the Abelian internal symmetry,
the maximal finite-dimensional subalgebra of $l^g$ contains
the superalgebra $osp(2,2)\oplus osp(2,2)$
spanned by generators $\Pi_{\pm}T^A$, where
\be
\label{Pi}
    \Pi_{\pm}=\frac{1\pm\psi_1}2
\ee
and $T^A=\{T_{\ga\gb}\,,\,Q_\ga^{(1)}\,,\,Q_\ga^{(2)}\,,\,J \}$ with
\be
\label{N2}
  T_{\ga\gb} = \frac1{4i}\{\ty_\ga,\ty_\gb \}_*\,,\quad
  Q_\ga^{(1)} = \ty_\ga     \,,\quad
  Q_\ga^{(2)} = \ty_\ga k   \,,\quad
  J = k+\nu \,.
\ee
The fact \cite{BWV} that these generators form $osp(2,2)$
is a simple consequence of the properties of the deformed oscillator
algebra (\ref{y mod}). Thus, the system (\ref{WW})-(\ref{SB})
possesses $N=2$ global SUSY for arbitrary $\nu$.

Now let us consider the extended systems. We start with the case
$\gb =1$. One can see that only the case $n=m$ corresponds to
the supersymmetric theory. At $n\neq m$, the algebras $hu_1(n,m|2)$
do not contain the AdS SUSY subalgebras.
Indeed, the generators of $osp(2,2)$ (\ref{N2})
can be embedded into these algebras as
$T_{\ga\gb}\otimes I_M\,,\;J\otimes I_M$ and
$Q_\ga^{(1)}\otimes Y_M\,,\;Q_\ga^{(2)}\otimes Y_M$, where
$$
    \begin{array}{rcl}
              Y_M & = & \left( \begin{array}{rcl}
                                     0   &    A \\
                                     B   &    0
                               \end{array}
                        \right)  \,,\qquad
    \end{array}
    Y_M^2 = I_M \,.
$$
This is possible only at $n=m$.

The case $\gb =0$ turns out to be supersymmetric at any $n$ and $m$
because the generators of $osp(2,2)$ are embedded into
 $hu_0(n,m|2)$ as $T^A\otimes I_M$.

To find the maximal $N=2$ supersymmetric
finite-dimensional global symmetry subalgebras
one should single out those Yang-Mills symmetries of sect.~\ref{YM}
that commute with $osp(2,2)\oplus osp(2,2)$.
For the case $\gb = 1$ and $n=m$, this is $u(n)\oplus u(n)$,
the diagonal subalgebra spanned by the $\psi_1$ - dependent matrices
(the doubling is due to $\psi_1$).
The maximal finite-dimensional subalgebra is
therefore $osp(2,2)\oplus osp(2,2)\oplus u(n) \oplus u(n)$.

In the case $\gb =0$ with arbitrary $n,m$,
the block-diagonal constant matrices commute with
$osp(2,2)\oplus osp(2,2)$ and, therefore, again
the maximal finite-dimensional
subalgebras are  $osp(2,2)\oplus osp(2,2)\oplus u(n)\oplus u(m)$.

Now let us analyze how maximal finite-dimensional global symmetries
act on the matter fields. We start with the Abelian case.
As we learned from sect.~\ref{LinEq}, the equation of motion for
the dynamical components $C^{dyn}(\ty;k,\psi_1|x)\,\psi_2$
in (\ref{decomp}) leads to the four infinite chains
(\ref{chainbos+-}) for the bosonic components and four
infinite chains (\ref{chainferm+-}) for the fermionic components.
Each chain realizes some infinite-dimensional representations of the
AdS algebra $o(2,2)=o(2,1)\oplus o(2,1)$ and describes a single particle.
With respect to $N=2$ SUSY $osp(2,2)\oplus osp(2,2)$,
all bosonic and fermionic components of $C^{dyn}$
belong to the same supermultiplet. This follows from the
following expansion of $C^{dyn}(\ty;k,\psi_1)$,
\be
\label{Cd ex}
   C^{dyn}(\ty;k,\psi_1)=[C^0_+(\ty)+C^0_-(\ty)]+
      [C^1_+(\ty)+C^1_-(\ty)]\psi_1 \,,
\ee
where in accordance with (\ref{pm})
$C^{0,1}_{\pm}=P_{\pm}C^{0,1}$, $P_{\pm}=\frac{1\pm k}2 $,
and from the realization (\ref{N2}) of the generators of
$osp(2,2)\oplus osp(2,2)$.
Thus we arrive at the $N=2$ hypermultiplet
constituted by 4 scalar and 4 spinor fields,
\be
\label{hypm}
   \left\{ \; C^0_+(x)\,,\,C^0_-(x)\,,\, C^1_+(x)\,,\,C^1_-(x)\,,\;
   C^0_{+\ga}(x)\,,\,C^0_{-\ga}(x)\,,\, C^1_{+\ga}(x)\,,\,C^1_{-\ga}(x)\;
     \right\}\,.
\ee
The masses are given by the formulae (\ref{M}) and (\ref{M f}) for scalars
and spinors respectively (the signs $\pm$ in (\ref{M}) and (\ref{M f})
correspond to those in (\ref{hypm})). The doubling of fields of
the same mass is due to the presence of $\psi_1$ and is typical for
$N=2$ hypermultiplets. In the complexified system the irreducible
subsystems are singled out by the conditions
$\Pi^\pm C^\pm = C^\pm$ with the projector (\ref{Pi}).
However, because the matter fields carry the factor of $\psi_2$
according to (\ref{decomp}),
$C^+$ and $ C^-$ turn out to be conjugated to each other
according to (\ref{hermit}). Another way to see this complex structure is
to observe that the similarity transformation
$C\rightarrow e^{i\alpha \psi_1 }C e^{-i\alpha \psi_1 }$
is equivalent to $C^\pm \rightarrow e^{\pm 2i\alpha }C^\pm $.

A generalization to the non-Abelian case is straightforward.
For the case $hu_{0}(n,m|2)$ the matter fields arrange into
$2( n\otimes \bar{m} \oplus m\otimes \bar{n} )$ representation
\footnote{In this paper we use the notations of type
$n\otimes \bar{m} \oplus m\otimes \bar{n}$ to denote a
complex field $\phi_i{}^{j^{\prime}}$
($i=1,\ldots,n,\quad j^{\prime}=1,\ldots,m$)
and its complex conjugate $\bar{\phi}_{j^{\prime}}{}^i$.}
of the global inner symmetry algebra $u(n)\oplus u(m)$.
Similarly, in the supersymmetric case $hu_{1}(n,n|2)$
the matter fields belong to the
$2(n\otimes \bar{n}^\prime \oplus \bar{n}\otimes n^\prime)$ representation
of the global inner symmetry algebra $u(n)\oplus u^\prime(n)$
(here primed indices are used to distinguish
between the two different subalgebras).

\subsection{$N=1$ SUSY}\label{N=1}

As shown in sect.~\ref{YM}, there exist $N=1$ supersymmetric
truncations of the system (\ref{WW})-(\ref{SB}), based on
antiautomorphisms. To study which of finite-dimensional global
symmetries survive in this case, let us turn to the truncation conditions
(\ref{trunc}) based on the symmetry (\ref{ruleW}) in the gauge field sector,
starting with the Abelian internal symmetry. Using the definition of
the antiautomorphism $\gs$ (\ref{sigma}), one finds that
this truncation eliminates the generators
$Q_\ga^{(2)}$ and $J$ from the set of $osp(2,2)$ generators (\ref{N2}),
while the generators $T_{\ga\gb}$ and $Q_\ga^{(1)}$ survive.
Thereby, this truncation breaks $N=2$ SUSY down to $N=1$ SUSY
$osp(1,2)\oplus osp(1,2)$ with the generators
\be
\label{N1}
   T_{\pm,\,\ga\gb} = \frac1{4i}\Pi_\pm
      \{\ty_\ga,\ty_\gb \} \,,\quad
   Q_{\pm,\,\ga} = \Pi_\pm  \ty_\ga     \,.
\ee

The conditions (\ref{trunc}), (\ref{ruleB}) eliminate from the expansion
(\ref{Cd ex}) the $\psi_1$-dependent components of the scalars $C_{\pm}$
and the fermionic components that can
depend on $k$ and $\psi_1$ only via $k\psi_1$. As a result, the truncated
$N=1$ matter
supermultiplet contains 2 scalars and 2 spinors,
\be
\label{supm}
   \left\{ \; C^0_+(x)\,,\; C^0_-(x)\,,\;
    C^0_{1\,\ga}(x) \,,\; C^1_{0\,\ga}(x)  \; \right\}
\ee
(with the convention $C(k) = C_0 + C_1\,k$)
with the masses (\ref{M}) and (\ref{M f}) for bosons and fermions
respectively.

The extended systems (\ref{WW})-(\ref{SB}) in this case have
the global symmetries $ho_{\gb}(n,m | 2)$ and $husp_{\gb}(n,m | 2)$
with the orthogonal and symplectic algebras in the Yang-Mills sector.
At $\gb =1$ and $n=m$, the maximal finite-dimensional global symmetry
subalgebras are
$osp(1,2)\oplus osp(1,2) \oplus o(n)\oplus o(n)$ or
$osp(1,2)\oplus osp(1,2) \oplus usp(n)\oplus usp(n)$,
depending on a sign in (\ref{specH})
(the doubling is again due to $\psi_1$).
At $\gb = 0$ ($\ga = 1$) and arbitrary $n, m$,
the maximal finite-dimensional subalgebra is either
$osp(1,2)\oplus osp(1,2) \oplus o(n) \oplus o(m)$ or
$osp(1,2)\oplus osp(1,2) \oplus usp(n)\oplus usp(m)$.

To find out what $N=1$ matter multiplets survive in the non-Abelian case
one should analyze the truncation conditions (\ref{trunc}), (\ref{ruleB})
in the sector of matter fields.
Let us start with the case $\ga =0$, $\gb =1$, and $n=m$,
which corresponds to the internal symmetries with diagonal $o(n)\oplus o(n)$
and $usp(n)\oplus usp(n)$. The dynamical matter fields $C$ (\ref{Cd ex})
are arranged into $2n\times 2n$ matrix
\be
\label{block}
\left( \begin{array}{ccc}
                    C_1     &   C_{1\,\ga}   \\
                    C_{2\,\ga}   &   C_2
             \end{array}
     \right) \,,
\ee
with bosons in the diagonal $n\times n$ blocks $C_1, C_2$, and fermions
in the off-diagonal $n\times n$ blocks $C_{1\,\ga}, C_{2\,\ga}$.
All components depend on $\psi_1$ and are proportional to $\psi_2$.
The dependence on $k$ is eliminated by the projection condition (\ref{proj}).
Then, the truncation condition for the
dynamical scalar fields
$C^{diag}(\psi_1 ) = C_1 (\psi_1) \oplus C_2 (\psi_1)$
reads
$$
    HC^{diag}(\psi_1) H^{-1} = (C^{diag} (-\psi_1 ))^T \,.
$$
In terms of matrix elements of the diagonal blocks this yields
taking into account (\ref{specH})
\be
\label{Hcomp}
    H^{ij}C_j{}^k (\psi_1 ) = H^{lk}C_l{}^i (-\psi_1 )
    = \pm H^{kl}C_l{}^i (-\psi_1 ) \,,\qquad
    i,j = 1,\ldots,n \quad \mbox{or} \quad
    i,j = n+1,\ldots,2n \,.
\ee
Denoting $(C^{diag})^{ik}=H^{ij}(C^{diag})_j{}^k$ we get
$(C^{diag}(\psi_1) )^{ik} = \pm (C^{diag}(-\psi_1))^{ki}$,
i.e. the components singled out by the projectors
(\ref{Pi}) turn out to be related to each other.

Taking into account that all dynamical fields are proportional to
$\psi_2$ and that the components in the direct sums
$o(n)\oplus o^\prime(n)$ and
$usp(n)\oplus usp^\prime(n)$ are singled out by the projector (\ref{Pi}),
one concludes that the physical bosons in each of the diagonal blocks
in (\ref{block})
belong to the representation
$n\otimes n^\prime$, so that the full set of the independent bosonic fields
belongs to the representation $2(n\otimes n^\prime)$ of the internal symmetry.
Analogous conditions for the dynamical spinors $C_\ga (\psi_1 ) =
C_{1\,\ga}(\psi_1 )\oplus C_{2\,\ga}(\psi_1 )$ read
\be
    C_{\ga}^{i'k''} (\psi_1) = \mp
    C_{\ga}^{k''i'} (-\psi_1) \,,\quad  i'= 1,\ldots,n \,,\quad k''
    = n+1,\ldots,2n  \,.
\ee
As a result we find that the fermions
also belong to the representation $2(n\otimes n^\prime)$.

In the case $\ga =1$ ($\gb =1$, $n=m$) the similar analysis
(with the symmetry properties of bosons and fermions interchanged)
leads to the same result with the matter fields in
the representation $2(n\otimes n^\prime)$ of the internal symmetries
$o(n)\oplus o^\prime (n)$ and $usp(n)\oplus usp^\prime (n)$.

The case $\ga =1$, $\gb =0$ with arbitrary $n$ and $m$ can be considered
analogously and corresponds to the representations
$2(n\otimes m)$ of the internal symmetries $o(n)\oplus o(m)$
and $usp(n)\oplus usp(m)$.

\subsection{Non-supersymmetric Case ($N=0$)}

The system (\ref{WW})-(\ref{SB}) with the global symmetries
$hu_\gb(n,m|2)$, $ho_\gb(n,m | 2)$, and \\ $husp_\gb(n,m | 2)$
at $\gb =1$ and $n\neq m$ do not possess global supersymmetries.
In this case one can consider only the $o(2,2)\oplus o(2,2)$ multiplets
in representations of inner symmetries.

In the systems based on $hu_{01}(n,m|4)$ with arbitrary $n$ and $m$,
bosons belong to the representation
$(n\otimes \bar{n}^\prime) \oplus (m\otimes \bar{m}^\prime)
\oplus (\bar{n} \otimes n^\prime) \oplus (\bar{m} \otimes m^\prime)$,
while fermions belong to the representation
$(n\otimes \bar{m}^\prime) \oplus (m\otimes \bar{n}^\prime)
\oplus (\bar{m} \otimes n^\prime) \oplus (\bar{n} \otimes m^\prime)$
of the internal symmetry
$u(n)\oplus u(m) \oplus u^\prime(n)\oplus u^\prime(m)$.

In the systems based on $ho_{01}(n,m|4)$ and $husp_{01}(n,m|4)$
bosons belong to the representation
$(n\otimes n^\prime)\oplus (m\otimes m^\prime)$, while fermions
belong to the representation
$(n\otimes m^\prime)\oplus (m\otimes n^\prime)$ of the internal symmetries
$o(n)\oplus o(m) \oplus o^\prime(n)\oplus o^\prime(m)$ or
$usp(n)\oplus usp(m) \oplus usp^\prime(n)\oplus usp^\prime(m)$.

In the case $\ga =1$ and $\gb = 1$, representations are
of the same structure, but
with the symmetry properties of bosons and
fermions interchanged.

Analogously, at $\gb =0$ ($\ga =1$) one can consider the
$o(2,2)\oplus o(2,2)$ multiplets in representations of broader inner
symmetries, if one does not require them to
commute with the SUSY subalgebras.
Then, in the systems based on $hu_{10}(n,m|4)$
bosons belong to the representation
$(n\otimes \bar{m}) \oplus (m\otimes \bar{n})
\oplus (n^\prime\otimes \bar{m}^\prime)
\oplus (m^\prime\otimes \bar{n}^\prime)$,
while fermions belong to the representation
$(n\otimes \bar{m}^\prime) \oplus (m\otimes \bar{n}^\prime)
\oplus (\bar{m} \otimes n^\prime) \oplus (\bar{n} \otimes m^\prime)$
of the internal symmetry
$u(n)\oplus u(m) \oplus u^\prime(n)\oplus u^\prime(m)$.
This is a manifestation of the important fact that the same
infinite-dimensional HS algebras can have different maximal
finite-dimensional subalgebras which contain the space-time
(AdS) symmetry algebra as a subalgebra.

In the systems based on $ho_{10}(n,m|4)$ and $husp_{10}(n,m|4)$
bosons belong to the representation
$(n\otimes m)\oplus (n^\prime\otimes m^\prime)$, while fermions
belong to the representation
$(n\otimes m^\prime)\oplus (m\otimes n^\prime)$
of the internal symmetries
$o(n)\oplus o(m) \oplus o^\prime(n)\oplus o^\prime(m)$ and
$usp(n)\oplus usp(m) \oplus usp^\prime(n)\oplus usp^\prime(m)$.

Finally, in the non-supersymmetric case it is of course possible
to truncate out fermions completely by imposing the condition
$f(A_i)=A_i$, $A_i = (W_\mu\,,\,B\,,\,S_\ga)$,
with $f$ (\ref{ap}).

\subsection{Massless Case and $N$-Extended SUSY}

It turns out that in the massless case $\nu=0$
the system (\ref{WW})-(\ref{SB}) admits an additional truncation
based on the automorphism $k\to -k$. The corresponding
involutive symmetry $\zeta$,
\be
\label{zeta}
   \zeta[W(k)] = W(-k) \,,\qquad
   \zeta[S_\ga(k)] = S_\ga(-k)  \,,\qquad
   \zeta[B(k)] = - B(-k) \,,
\ee
induces in the $\beta=0$ case
 the truncation conditions
$\zeta(A_i) = A_i$, $ A_i = (W_\mu\,,\,B\,,\,S_\ga) $.
It is important that at $\nu = 0$ the vacuum solutions (\ref{B_0}),
(\ref{S sym}), and (\ref{W_0}) are compatible with these conditions.

In the non-Abelian case with $\gb=1$ and $n=m$, one should consider
the automorphism $k\to -k$ combined with some involutive automorphism
$\chi$, $\chi(U) = - U$, to ensure that the projection conditions
(\ref{proj}) are invariant. One can define $\chi$ on $Mat_{2n}$
as
\be
\label{chi}
    \chi(M) = YMY^{-1} \,,\quad
    \begin{array}{rcl}
          Y & = & \left( \begin{array}{ccc}
                                0    &    I_n \\
                               I_n   &     0
                         \end{array}
                  \right)  \,.
    \end{array}
\ee
As a result, a generalization of the truncation induced by (\ref{zeta})
to all supersymmetric cases is the truncation by $\zeta\chi^\gb$.

The truncation based on $\zeta$ reduces our system
to that proposed in \cite{Eq}.
The reduced system possesses $N=1$ SUSY
$osp(1,2)\oplus osp(1,2)$ with the generators (\ref{N1}).
The set of fields obtained from (\ref{hypm}) with the aid of this
truncation turns out to be reducible. One can truncate the system
further with the aid of the symmetry (\ref{ruleW})-(\ref{ruleS})
which preserves $N=1$ SUSY. As a result, in the Abelian case we
arrive at the following $N=1$ massless supermultiplet
\be
\label{supm N1}
   \{ \; C^0_1(x) \,,\; C^0_{1\,\ga}(x) \;\} \,.
\ee
In the non-Abelian case, we arrive at the massless
$N=1$ supermultiplets in the representations
$n\otimes n^\prime$ ($\gb=1$, $n=m$) or $n\otimes m$ ($\gb=0$, $\ga=1$)
of the corresponding internal symmetries
of subsect.~\ref{N=1}. (Note that the
non-Abelian truncation based on the symmetry $\zeta\chi^\gb$
does not affect the inner symmetries.)

In the massless case $\nu=0$ there exists an interesting
alternative truncation
based on the symmetry (\ref{zeta}), which preserves
$N=2$ SUSY. Consider the combination $\zeta\chi^\gb\eta$ of
the symmetries (\ref{zeta}), (\ref{chi}) and the symmetry
(\ref{ruleW})-(\ref{ruleS}) based on the antiautomorphism $\gs$ of
subsect.~\ref{Anti}.
In the Abelian case, the corresponding truncation conditions
reduce the $N=2$ hypermultiplet (\ref{hypm}) to the following
$N=2$ massless supermultiplet,
\be
\label{supm N2}
   \left\{ \; C^0_1(x)\,,\, C^1_0(x)\,,\,
   C^0_{0\ga}(x)\,,\, C^0_{1\ga}(x) \;  \right\}  \,.
\ee
In the non-Abelian case, we arrive at the $N=2$ massless supermultiplet
either in the representations $2(n\otimes n^\prime)$ of the internal
symmetries $o(n)\oplus o^\prime(n)$ and $usp(n)\oplus usp^\prime(n)$
($\gb =1$, $n=m$), or in the representations $2(n\otimes m)$ of
$o(n)\oplus o(m)$ and $usp(n)\oplus usp(m)$ ($\gb =0$, arbitrary $n,m$).
This additional reduction compatible with $N=2$ SUSY is a manifestation
of the well-known shortening of massless supermultiplets.

Finally let us discuss a possibility to have extended supersymmetry
with $N>2$. In \cite{OP1}, it was shown that there is a simple way to
incorporate $N$-extended superalgebras $osp(N,2m)$
($m=1$ for the $d=3$ case under consideration) by supplementing
the bosonic generating elements of the Heisenberg algebra
$y_\alpha$ with the Clifford elements $\phi^i$ ($i=1,\ldots, N$). In this
approach the Clifford algebra $C_N$ is a particular case of the matrix
algebra $Mat_{2^{\frac{N}{2}}}$ (for $N$ even),
 while the generators of the $osp(N,2m)$ are
 realized in terms of the bilinears
\be
\label{genext}
T_{\alpha\beta} = \{y_\alpha , y_\beta \} \,, \qquad
   Q^i_\alpha = y_\alpha \phi^i \,,\qquad M^{ij} = [\phi^i ,\phi^j ]  \,,
\ee
provided that
\be [y_\alpha, y_\beta ] = 2i \epsilon_{\alpha\beta } \,,\qquad
    \{\phi^i, \phi^j \} = 2\delta^{ij} \,.
\ee
It is this realization of the $d4$ $N=8$ extended supersymmetry which was
recently used in \cite{sez} to discuss the $N=8$ version of the $d=4$ HS model.

Now we observe that this construction is not working for the deformed
oscillators               because the generators
(\ref{genext}) do not form a closed algebra if $y_\alpha$
is replaced by $\hat{y}_\alpha$ (\ref{y mod})
with $\nu\neq 0$.
This result can be explained as follows. For $\nu \neq 0$ the mass of
the matter supermultiplet is non-vanishing and the spin range within a
supermultiplet increases with $N$. Since
massive fields of spins greater than 1/2 are not included in our model,
$N>2$ extended supersymmetry cannot be realized.
In the massless case, however, one can realize higher supersymmetries within
only scalar and spinor fields \cite{dwtol} due to trivialization of
the notion of spin for the $d=3$ massless case (i.e. trivialization of
the $d=3$ massless little group).

Thus we conclude that the model under consideration admits $N > 2$
extended supersymmetry only for the massless vacuum $\nu = 0$.
The conclusion that arbitrary high $N$ is allowed differs from the
conclusions of \cite{dwtol} because our model is more general due
to the presence of the HS gauge interactions.
It is worth to mention that $d=3$ massless fields can be interpreted
\cite{SezNic} as $d=4$ singletons.

\section*{Conclusion}

The $d3$ model analyzed in this paper is shown to describe
HS gauge interactions of massive $N=2$ hypermultiplets.
The parameter of mass of the matter fields appears as a vacuum
expectation value of a certain auxiliary scalar field.
The model admits a generalization with classical Yang-Mills groups
which are of the unitary type for the $N=2$ supersymmetric
case and either orthogonal or symplectic for the $N=1$ reductions,
in striking parallelism with superstring theory.
Moreover, similarly to the previously obtained results for $d4$
HS models \cite{Ann,KV1}, the $N=1$ reductions are obtained with
the aid of a certain antiautomorphism of the Heisenberg-Weyl algebra,
which is a counterpart of the twist (orientation) operator in
string theory.

An important result of the paper, which is expected to have
implications for the theory of HS gauge fields in various dimensions
(and, hopefully, also for its lower-spin reductions like
supergravity), is that the proposed dynamical equations
admit an integrating flow which relates solutions of the non-linear
system to those of the linearized system and is in many respects
similar to the B\"acklund transformations in integrable systems and to
the Nicolai map \cite{Nic} in supersymmetric models. Apart from
the fact that this integrating flow allows one to build
constructively perturbative solutions of the model, it raises
an important question of the proper definition of the concept of
locality in the models with a dimensionful parameter like
a cosmological constant, which we hope to discuss in a future publication.
As we demonstrate in this paper, a similar phenomenon takes place in
ordinary $d3$ gravity with the cosmological term, where
it is possible to perform a weakly local (i.e. containing
non-localities that can be expanded in infinite series of higher
derivatives) field redefinition
which compensates the stress energy tensor on the right hand side
of the equations of motion
thus reducing the problem to the vacuum case with zero-curvature
equations. An important difference, however, is that in the case of the HS
models we are able to write down in a very simple and explicit form the
integrating flow that governs such a field redefinition in all orders in
interactions. A related interesting question consists in
the generalization of the cohomological analysis of the stress-energy
tensor,
performed in Appendix C of this paper, to all conserved HS currents.
One can expect that, similarly to the case of gravity, all HS currents
belong to the trivial cohomology class with respect to the weakly local
transformations, but have to belong to a non-trivial cohomology class
with respect to local transformations.

Another interesting related topic is a proper definition of locality in
the $d4$ HS gauge theories.
In the context of the results of this paper one can speculate that
the fact that $d4$ HS gauge theories require non-zero cosmological
constant \cite{FV1} might be a signal of some sort of non-locality of
the HS gauge theories beyond the cubic order
(in the cubic order the number of derivatives
of some field of a given spin $s$ in the interaction part of the action
is bounded by $s$ \cite{FV1}). This hypothesis is very interesting
in the context of the applications to $M$ theory.

An intriguing problem for the future is to analyze
in detail what happens
at the special points $\nu = 2l+1$, $l\in {\bf Z}$.
In \cite{BPV}, it is shown that at these points the free field
equations effectively start from the some higher Lorentz multispinors
rather than from a scalar. The same singular points appear in the
vacuum solution (\ref{S sym})
(see also Appendix B). Remarkably, the
values $\nu =1, 3$ correspond to the case of $d3$ Maxwell
electrodynamics \cite{BPV}.
The physical meaning of the higher singular values of $\nu$ is
not yet clear. The mathematical interpretation is very simple however.
These special values correspond to all those values of the Casimir
operators of the AdS algebra $o(2,2)\sim sp(2)\oplus sp(2)$ which correspond
to its finite-dimensional representations. In other words, the situation
becomes special when the relevant representations of the AdS algebra
admit singular vectors while the infinite-dimensional HS algebra acquires
ideals.

\section*{Acknowledgments}

Authors are grateful to R.~R.~Metsaev for a useful comment.
This research was supported in part by INTAS, Grants No.96-0538, No.96-0308,
and by the RFBR Grant No.96-01-01144.
S.~P. acknowledges a partial support from the Landau Scholarship
Foundation, Forschungszentrum J\"ulich.

\setcounter{section}{0}
\def\thesection{Appendix \Alph{section}.}
\def\theequation{\Alph{section}.\arabic{equation}}

\section{Regularity}\label{Reg}

In this appendix, we define a class of functions
in the auxiliary spinor spaces, used throughout the paper.
This regularity class generalizes that introduced in \cite{Prop}.

\noindent

{\it Definition.}
A function $f(z,y;k,\psi_{1,2})$ is called regular if it can be expanded
into a finite sum of some functions $g$ of the form
\be
\label{A1}
g(z,y;k,\psi_{1,2})=P(z,y;k,\psi_{1,2})\;
      \int_{M^n}\,d^n t\,\rho(t)\,\exp[i\phi(t)(zy)] \,,
\ee
where the integration is carried out over some compact domain
$M^n \subset R^n$
with the coordinates $t_i$ ($i=1, \ldots, n$), $P(z,y;k,\psi_{1,2})$ is an
arbitrary polynomial of $z,y,k$ and $\psi_{1,2}$, $\phi(t)$ is an arbitrary
polynomial function of $t_i$, while $\rho(t)$ is some absolutely integrable
function on $M^n$.
\\
\\
{\it Comment 1}. This definition contains individual exponentials
with arbitrary polynomial pre-exponential factors via formulae like
$ \exp  a(zy) = \int_0^1 dt \frac{\partial}{\partial t}\left[t \exp[ta(zy)]
\right] \,.$
Equivalently one can allow $\rho(t)$ to contain an arbitrary finite
number of $\gd$-functions and their derivatives.
\\
\\
{\it Theorem.} Given regular functions $g_1 (z,y;k,\psi_{1,2})$ and
$g_2 (z,y;k,\psi_{1,2})$, their product (\ref{prod}) $(g_1 *g_2
)(z,y;k,\psi_{1,2})$ is some regular function.
\\
\\
{\it Proof.}
$$ g_1*g_2 = P_1(z,y)\int_{M_1} dt_1 \rho_1(t_1) \exp [i\phi_1(t_1)(zy)]
\quad *
\quad P_2(z,y)\int_{M_2} dt_2 \rho_2(t_2) \exp [i\phi_2(t_2)(zy)]=
$$ $$
   =\int_{M_1} dt_1 \rho_1(t_1) \exp [i\phi_1(t_1)(zy)] \int_{M_2} dt_2
    \rho_2(t_2) \exp [i\phi_2(t_2)(zy)] \times $$
$$ \times {1\over (2\pi)^2}
   \int\int du\,dv\,\exp\{i(uv)+i\phi_1(t_1)[(z-y)u]+i\phi_2(t_2)[(z+y)v]\}
       \,P_1(z+u,y+u) P_2(z-v,y+v)  \,.
$$
Inserting
\be
   P(z+a,y+b)=\left. \exp\left[a^\ga {\partial\over \partial z_1^\ga}+
      b^\ga {\partial\over \partial y_1^\ga}\right]
      P(z_1, y_1) \right|_{z_1 = z \atop y_1 = y}\,,
\ee
one gets
$$
   g_1*g_2 = \int_{M_1\times M_2} dt_1\,dt_2\,\rho_1(t_1)\rho_2(t_2)
   \exp\{ i[\phi_1(t_1) + \phi_2(t_2)](zy) \}  \;\times
$$
$$
   \times \left. {1\over (2\pi)^2}\int\int
          du\,dv\, \exp\left\{u^\ga\left[i\phi_1(t_1)(z-y)_\ga+
          {\partial\over \partial z_1^\ga}+
          {\partial\over \partial y_1^\ga}  \right]\right\}\,
          P_1(z_1, y_1) \right|_{z_1=z\atop y_1=y} \times
$$
$$
   \times \left. \exp\left\{iv^\ga\left[\phi_2(t_2)(z+y)_\ga+u_\ga-
          i\left({\partial\over \partial y_2^\ga}-
          {\partial\over \partial z_2^\ga} \right) \right]\right\}\,
          P_2(z_2, y_2) \right|_{z_2=z\atop y_2=y}=
$$
$$
   =\int_{M_1\times M_2} dt_1\,dt_2\,\rho_1(t_1)\rho_2(t_2) \;
      \exp\left\{i[\phi_1(t_1)+\phi_2(t_2)-2\phi_1(t_1)\phi_2(t_2)](zy)
        \right\} \times
$$
$$
   \times \exp\left\{-i \gep^{\ga\gb} \left({\partial\over \partial z_1^\ga}
      + {\partial\over \partial y_1^\ga}  \right)
      \left({\partial\over \partial z_2^\gb} -
      {\partial\over \partial y_2^\gb}  \right) \right\}
      P_1\left[z_1-\phi_2(t_2)(z+y)\,,\,y_1-\phi_2(t_2)(z+y) \right]\; \times
$$
$$
   \times P_2\left[z_2-\phi_1(t_1)(z-y)\,,\,y_2+\phi_1(t_1)(z-y) \right]
      \Bigg|_{z_1 = z_2 = z \atop y_1 = y_2 = y}  \,.  
$$
Since the product of the two compact domains $M_1 \subset R_n$ and
$ M_2 \subset R_m$ is a compact domain in $R_{n+m}$ and $P_1\,,\,P_2$
are some polynomials, one concludes that the latter expression is
a finite sum of some regular functions (\ref{A1}).
\\
\\
{\it Comment 2}. In this proof it is important that the determinant
of the quadratic form in the Gaussian integral is a constant independent
of the particular choice of
the functions $g_1$ and $g_2$. This property is a consequence of
the fact that we use the star product (\ref{prod}) and might
not be true for other star-product formulae. As an example,
the reader can check that
$\exp\,[i(zy)] *\exp\,[i(zy)] $
does not exist for  the star product corresponding to the Weyl ordering.

{}From the theorem above it follows that every finite-order perturbative
computation that starts with some regular
generating functions $W,B$, and $S$ is well defined.
In particular the vacuum solutions $S_{0\ga}$
 and $\tilde y_\ga$ belong to the regularity class.

\section{Construction of Vacuum Solutions}\label{GV}

The problem is to find elements $\ty_\ga$ and
$S_{0\ga}$ such that
\be
\label{y1}
   [\ty_\ga,\ty_\gb]_* =
      2i\gep_{\ga\gb}(1+\nu k)\,,\quad
      \ty_\ga k = - k \ty_\ga \,,
\ee
\be
\label{S1}
   [S_{0\ga},S_{0\gb}]_* =
      -2i\gep_{\ga\gb}(1+\nu K) \,,\quad
      S_{0\ga} * K = - K * S_{0\ga} \,,
\ee
\be
\label{Sy}
   [S_{0\ga},\ty_\gb]_* =0 \,.
\ee
Let us start with $\ty_\ga$. A Lorentz covariant Ansatz is
\be
\label{ans ty}
   \ty_\ga =\int^1_{-1} ds
   [ (y_\ga +z_\ga ) n(s,k) + (y_\ga - z_\ga ) m(s,k) ]
   \exp \left[ \frac{i}{2} (s+1) (z_\ga y^\ga) \right] \,.
\ee
A direct computation shows that
\bee
\label{yy}
   \ty_\ga * \ty^\ga & = &
   \int ds ds^\prime \; \theta (1+s) \theta (1-s)
   \theta (1+s^\prime ) \theta (1-s^\prime )
   \exp \left[\frac{i}{2} (1-ss^\prime ) (z_\ga y^\ga) \right]\nonumber\\
   & \times & [(8i+2ss^\prime (z_\ga y^\ga )) n(s,-k)m(s^\prime, k)
   -2 (z_\ga y^\ga ) m(s,-k) n(s^\prime, k)]  \,.
\eee
Let us require the following condition to be true,
\bee
\label{cond1}
   & \displaystyle {\int ds ds^\prime \; \theta (1+s) \theta (1-s)
   \theta (1+s^\prime ) \theta (1-s^\prime )  \gd (t-ss^\prime )
   n(s,-k) m(s^\prime ,k) }  & \nonumber \\
   & \displaystyle { {} = {1\over 8} (1+t)[\gd(t-1)+\gd(t+1)+\nu k] \,. }&
\eee
Inserting this into (\ref{yy}) one finds that
\be
\label{y2}
   \ty_\ga * \ty^\ga = 2i(1+\nu k)   \,,
\ee
which is an equivalent form of (\ref{y1}).

Thus, every solution of (\ref{cond1}) gives rise to some
$\ty_\ga$ (\ref{y1}). Let us now construct a class of solutions
of (\ref{cond1}) with the aid of the following Ansatz
\be
\label{nm}
   n(s,k) = {1\over 2} (1+\gga s)p(s,k)  \,,\qquad
   m(s,k) = {1\over 2} (1+\gga s)q(s,k)  \,
\ee
with
\be
\label{even}
   \gga = \pm 1 \,,\qquad
   p(s,k) = p(-s, k) \,,\qquad
   q(s,k) = q(-s, k) \,.
\ee
These symmetry properties prove that
\be
\int_{-1}^1 ds \int_{-1}^1 ds^\prime \;
\gd (t-ss^\prime )(s + s^\prime) p(s)q(s^\prime)=0\,.
\ee
As a result (\ref{cond1}) reduces to the form
\be
\label{cond2}
   (\;p(-k) \circ q(k) \;) (t)=
   {1\over 2} [\gd (t-1) + \gd (t+1)+\nu k]  \,,
\ee
where a new product $\circ$ is defined by the relation
\be
\label{circ}
   (p\circ q) (t) = \int ds ds^\prime \; \theta (1+s) \theta (1-s)
   \theta (1+s^\prime ) \theta (1-s^\prime ) \gd (t-ss^\prime )
   p(s)q(s^\prime) \,.
\ee
It is easy to see that this product law is commutative and
associative.

To solve (\ref{cond2}) it is convenient to use the following trick.
Consider the expansion
\be
\label{pI}
   p(t) = \sum_{n=0}^\infty p_n I_n (t) \,,
\ee
where the basis functions $I_n$ are defined by the multiple integrals
\be
   I_n (t) = \int d^n s \theta(1 + s_1) \ldots \theta(1 + s_n)
   \theta (1-s_1)\ldots \theta(1 - s_n )  \gd (t-s_1 s_2 \ldots s_n )
\ee
for $n>0$, and
\be
\label{I_0}
   I_0 = \frac{1}{2} \left[\gd (t-1) +\gd (t+1) \right]  \,.
\ee
It is easy to evaluate the basis integrals $I_n$,
\be
\label{I}
   I_n = \frac{(-\mbox{ln} \, t^2 )^{n-1}}{(n-1)!} \,, \qquad n>0 \,.
\ee
One can check that so defined functions $I_n$ have the following
basic property,
\be
\label{basic}
   I_n \circ I_m = I_{n+m}\,\quad \forall n \ge 0\,.
\ee
Therefore, given $p(s)$ (\ref{pI}) we can define its symbol $\tilde{p}(x)$,
\be
\label{ssym}
   \tilde{p}(x) = \sum_{n=0}^\infty p_n x^n \,,
\ee
an ordinary function of the variable $x$.
{}From (\ref{basic}) it follows then that
\be
   \widetilde{(p\circ q)}(x)= \tp (x) \tq (x)\,.
\ee

The equation (\ref{cond2}) rewritten in terms of symbols
has a very simple form
\be
\label{symeq}
   \tp (x,-k) \; \tq (x,k) = 1 + {1\over 2} \nu k x
\ee
and admits a class of solutions parametrized by an arbitrary function.
This ambiguity takes its origin in the gauge ambiguity (\ref{S_1}).
A solution of particular interest is the one that transforms
properly under
the involution (\ref{inv}) and
the antiautomorphism (\ref{sigma}),
$\sigma (\tilde{y}_\alpha )=i\tilde{y}_\alpha  $ and
$\tilde{y}^\dagger_\alpha =\tilde{y}_\alpha$.
This is the case provided that
\be
   n(s,-k) = m(s,k) \,,\qquad \bar{n}(s,k) = n(s,k) \,, \qquad
   \bar{m}(s,k) = m(s,k) \,.
\ee
According to (\ref{nm}), this implies that
 $p$ and $q$ are real and
\be
\label{symsol}
   p(s,-k) = q(s,k) \,.
\ee
Now we see that there is only one (real) solution of (\ref{symeq}),
satisfying (\ref{symsol}),
\be
\label{qp sym}
   \tq_{sym} (x,k) = \sqrt{1 + {1\over 2}\nu k x } \,, \qquad
   \tp_{sym} (x,k) = \sqrt{1 - {1\over 2}\nu k x } \,.
\ee
Thus, the symmetry requirement selects
a vacuum solution uniquely. Relaxing this symmetry requirement
one can find many solutions. The simplest ones are
\be
\label{+}
   \tq = 1 \,, \qquad  \tp = 1 - {1\over 2}\nu k x   \,,
\ee
or
\be
\label{-}
   \tq = 1 + {1\over 2}\nu k x  \,, \qquad \tp = 1   \,.
\ee

One can consider a one-parametric interpolating class
of solutions $\{ \ty_\ga^\tau \}$,
\be
\label{qp}
   \tq_\tau (x,k) = \left(1 + {1\over 2} \nu k x \right)^\tau \,, \qquad
   \tp_\tau (x,k) = \left(1 - {1\over 2} \nu k x \right)^{1-\tau} \,.
\ee
Once the symbols $\tp$ and $\tq$ are known, one can
compute the expansion coefficients in (\ref{ssym})
and then reconstruct the original functions $p(s,k)$ and
$q(s,k)$ via (\ref{pI}), (\ref{I_0}), and (\ref{I}).
In what follows, we will set $\gga=-1$ to have solutions that
start with $y_\ga$ in the zero order in $\nu$
\footnote{In the case $\gamma = 1$ a solution contains an additional
exponential factor, but it can be shown to be equivalent to that with
$\gamma =-1$ by an automorphism.}.

The polynomial solutions (\ref{+}), (\ref{-}), which correspond
to $\tau=0$ and $\tau=1$, give rise, respectively, to the
solutions $\ty_\ga^+$ and $\ty_\ga^-$ (\ref{tilde y}).
The symmetric solution (\ref{qp sym}) corresponds to $\tau={1\over 2}$
and gives rise to $\ty^{sym}_\ga$ (\ref{y sym}).

Expanding (\ref{qp}) in power series and using (\ref{I_0}) and (\ref{I})
one finds
\be
   q_\tau (s,k) = {1\over 2} \left[\gd (s-1) + \gd(s+1) +
   k \nu \tau \; \Phi (1-\tau\,, \; 2\,; \; \mbox{ln} \,|s|^{\nu k}) \right]
\ee
and
\be
   p_\tau (s,k) = {1\over 2} \left[\gd (s-1) + \gd(s+1) -
   k \nu (1-\tau) \; \Phi (\tau\,, \; 2\,; \; -\mbox{ln} \,|s|^{\nu k})
   \right] \,.
\ee
Here $\Phi (a,c;x )$ is the degenerate hypergeometric function
defined by the power series
\be
   \Phi (a,c;x ) = 1+\frac{ax}{c1!}+\frac{a(a+1)x^2}{c(c+1)2!}+ \ldots \,.
\ee
At $0< a < c$ it admits a useful integral representation \cite{sprav}
\be
   \Phi (a,c;x)=\frac{\Gamma(c)}{\Gamma(a)\Gamma(c-a)}\int_0^1 dt\;
   e^{tx} t^{a-1}(1-t)^{c-a-1}  \,,
\ee
which leads to
\be
   q_\tau (s,k) = {1\over 2} \left[\gd (s-1) + \gd (s+1) +
   k \; \frac{\nu \tau}{\Gamma(1-\tau)\Gamma(1+\tau)}
   \int_0^1 dt\; |s|^{t \nu k} t^{-\tau} (1-t)^\tau \right]  \,,
\ee
\be
   p_\tau (s,k) = {1\over 2} \left[\gd (s-1) + \gd (s+1)-
   k \; \frac{\nu (1-\tau)}{\Gamma(\tau)\Gamma(2-\tau)}
   \int_0^1 dt\; |s|^{-t \nu k} t^{\tau-1} (1-t)^{1-\tau} \right]  \,,
\ee
with $0< \tau < 1$.

{}From these formulae we see that the corresponding $\tilde{y}_\ga$
(\ref{ans ty}) is well defined at least in some
neighborhood of $\nu = 0$.
The case of $\tau = \frac{1}{2}$ is most interesting.
It is not difficult to see that the solution $\tilde{y}^{sym}$ makes sense
for $\nu < 1$ while some singularity can appear at $\nu = 1$. This
is just the closest to zero special value of $\nu$ discussed in
 sect.~\ref{OpReal}.
In fact, we  expect that our results can be extended
to arbitrary values of $\nu$ with some specificities
for $\nu = 2l+1$, $l\in {\bf Z}$. To achieve this one has to
analyze the problem more accurately at the singular point $t=0$ in
(\ref{I}).
We hope to discuss
this intriguing issue elsewhere.

To find $S_{0\ga}$ satisfying (\ref{S1}),
one can use the same Ansatz (\ref{ans ty})
with the substitutions  $k\to K = k e^{i(z_\ga y^\ga)}$,
$n(s,k)\to n(s,k)$,
$m(s,k)\to -m(s,k)$, $y_\ga \to \rho y_\ga$ and $z_\ga \to \rho z_\ga$,
\bee
\label{ans tS}
   S_{0\ga} &=& \rho \int^1_{-1} ds
   \left\{ (y_\ga +z_\ga ) \exp \left[ \frac{i}{2} (s+1) (z_\ga y^\ga)
   \right]  \;*\; n(s,K) \right. \\ \nonumber
   && {} \left.-(y_\ga - z_\ga )
   \exp \left[ \frac{i}{2}(s+1)(z_\ga y^\ga) \right] \;*\; m(s,K) \right\}
   \,.
\eee
Again, (\ref{S1}) is a consequence of (\ref{cond1}).

Further, it can be shown that the elements $\ty_\ga$ (\ref{ans ty})
and $S_{0\ga}$ (\ref{ans tS}) with the same functions
$n(s,k)$ and $m(s,k)$ commute to each other provided that (\ref{cond1})
is true. To prove this, it is convenient to use the projected
expressions
$[S_{0\ga}\,,\, \ty_\gb]_* *{1\over 2}(1\pm \exp(iz_\gga y^\gga))$.

To summarize, we arrive at the class of the solutions
$S^{\tau}_{0\ga}$ and $ \ty^{\tau}_\gb$
of (\ref{y1})-(\ref{Sy}).
The solutions ($S^+_{0\ga}$, $\ty^+_\gb$), ($S^-_{0\ga}$, $\ty^-_\gb$),
and ($S^{sym}_{0\ga}$, $\ty^{sym}_\gb$) correspond to the cases $\tau = 0$,
$\tau =1$, and $\tau = \frac{1}{2}$ respectively.

At the end of this Appendix, let us note that
according to Eqs. (\ref{I}), (\ref{ssym}),
the problem of reconstruction of $p(s)$ from its symbol
is equivalent to inverting the following remarkable mapping:
given $f(x) = \sum_{n=0}^\infty f_n x^n$,
$\tilde{f} = \sum_{n=0}^\infty \frac{1}{n!} f_n x^n$.
This mapping has a number of remarkable properties. In particular,
\be
   \widetilde{(f^\prime)} = x (\tilde{f})^{\prime\prime} +
     (\tilde{f})^{\prime}\,,\qquad
   \widetilde{(xf)} = \int_0^x dz \tilde{f}(z)\,.
\ee

\section{Example of Gravity}\label{Grav}

In this appendix, we use an example of gravity to illustrate
how non-local transformations allow one to integrate out
non-trivial currents (stress-tensor).

Consider some matter field $C$ interacting with the gravitational field
$W^{\ga\gb}=\go^{\ga\gb}+\gl\psi h^{\ga\gb}$
expanded near the AdS  vacuum solution
$W_0^{\ga\gb}=\go_0^{\ga\gb}+\gl\psi h_0^{\ga\gb}$
described in sect.~\ref{OpReal}.
Einstein equations with the cosmological term and a matter source in 2+1
dimensions can be written in the form
\be
\label{Ee1}
      R_{\ga\gb}=d\go_{\ga\gb}-\go_{\ga\gga}
      \wedge\go_\gb{}^\gga-\gl^2h_{\ga\gga}
      \wedge h_\gb{}^\gga={\cal T}_{\ga\gb}(C) \,,
\ee
\be
\label{Ee2}
    r_{\ga\gb}=dh_{\ga\gb}-\go_{\ga\gga}
      \wedge h_\gb{}^\gga-\go_{\gb\gga}
      \wedge h_\ga{}^\gga=0 \,,
\ee
or, equivalently,
\be
\label{Ee}
    dW_{\ga\gb}-W_{\ga\gga}\wedge W_\gb{}^\gga=
      {\cal T}_{\ga\gb}(C)  \,.
\ee
Here $R_{\ga\gb}$ and $r_{\ga\gb}$ are the curvature
and torsion 2-forms respectively.
The 2-form ${\cal T}_{\ga\gb}(C)$ arises from
the energy-momentum tensor $T_{\mu\nu}(C)$ in gravity,
when rewriting the Einstein equations
$R_{\mu\nu}-{1\over2} g_{\mu\nu} R = T_{\mu\nu}(C)$
(to simplify formulae we discard the coupling constant) in
the form (\ref{Ee1}), which is possible in $d=3$. Note that it is
assumed in (\ref{Ee}) that ${\cal T}_{\ga\gb}(C)$ is independent
of the Clifford
element $\psi$ ($\psi^2=1$), which distinguishes between $\go^{\ga\gb}$ and
$h^{\ga\gb}$, i.e. the torsion is zero.

Schematically, the mechanism is as follows.
Linearized Einstein equations can be cast into the form
(with appropriate gauge fixings)
\be
\label{lein}
   \left(L^C -\Lambda^2\right) h_{\mu\nu}=T_{\mu\nu}(C)  \,,
\ee
where $h_{\mu\nu}$ is the fluctuational part of the metric tensor,
$L^C$ is the linear operator of the equations of
the matter fields $L^C C =0$,
while $\Lambda = \ga \gl $ with some numerical coefficient
$\ga\neq 0$ which may depend on a particular choice of the matter
sources (e.g. on spin).
It is important that when the cosmological constant is non-vanishing,
the term with $\Lambda^2$ turns out to be non-vanishing too because
originally massless fields in the flat space acquire different mass-type
terms with the parameter of mass proportional to the inverse AdS radius
$m_s \sim \gga_s \gl$. We will call this phenomenon
spectra separation. It allows one to solve formally
(\ref{lein}) by a field redefinition
\be
\label{nonl h}
    h_{\mu\nu}'=h_{\mu\nu}-
     \left(L^C -\Lambda^2\right)^{-1} T_{\mu\nu}(C)=
     h_{\mu\nu}+{1\over \Lambda^2} \sum_{n=0}^\infty
        \left({1\over \Lambda^2}L^C\right)^n T_{\mu\nu}(C) \,.
\ee
Note that it is important here to use the operator $L^C$ to avoid
infinite resummations due to collecting similar terms of the type
$(L^C )^n C (L^C )^m C $ which would all be proportional to $C^2$
if $L^C ( C)\sim C \neq 0$. It is also clear
that a non-vanishing dimensionful constant, the cosmological constant,
plays crucial role in this analysis. A non-local character of the
transformation manifests itself in the appearance of  infinite series
in inverse cosmological constant (equivalently, the radius of the AdS space).

{}From the equation (\ref{Ee}) it is obvious that this analysis
is cohomological in nature: the form
${\cal T}_{\ga\gb}(C)$ is closed on-mass-shell (this is $d3$
stress-energy conservation law) but it turns out to be exact, once
one relaxes the locality condition.

To make the analysis above more explicit let us consider an example
of a massless scalar matter field, using the ``unfolded form''
(\ref{chainbos+-}) of its free equations of motion
$\Box C={3\over 2}\gl^2 C$ at $\nu=0$  (with
the choice ``+''),  
\be
\label{chainb}
      D_0^L C_{\ga(n)}=\frac i2 \psi \left[h_0^{\gb\gga}
      C_{\gb\gga\ga(n)}-\gl^2n(n-1)h_{0\,\ga\ga}
      C_{\ga(n-2)}\right] \,,
\ee
where $D_0^L$ is the background covariant differential (\ref{Lcov D0})
and the scalar component $C$ is supposed to be $\psi$-independent.
Taking into account (\ref{chainb}) one can write
${\cal T}_{\ga\gb}(C)$ in the form
\be
\label{calT}
   {\cal T}_{\ga\ga}(C)={1\over 4} h^\gga{}_\gb \wedge
      h^{\gb\gd} C_{\gga\ga} C_{\gd\ga}-
      {1\over 2}\mu^2 h_{\ga\gb} \wedge
      h^\gb{}_\ga C^2  \,,
\ee
where $\mu$ is the ``AdS mass'', $\mu^2=-{3\over 2}\gl^2$.
One can see that ${\cal T}_{\ga\ga}(C)$ is $\psi$-independent because
in accordance with (\ref{chainb})
$C_{\ga(n)} \sim \psi^{{n\over 2}}$.

The equation (\ref{Ee}) in the lowest order reads
\be
\label{lin E}
    D_0 W_{1\,\ga\ga}={\cal T}_{\ga\ga}(C)  \,,
\ee
where $D_0$ is the full background covariant differential,
\be
\label{cov D0}
    D_0 A_{\ga (n)}=dA_{\ga (n)}+nW_{0\ga}{}^\gga \wedge
       A_{\gga\ga(n-1)}  \,.
\ee

Suppose there exists such a field redefinition
\be
\label{shift W}
    W_{1\,\ga\gb}'=W_{1\,\ga\gb}-
      \psi U_{\ga\gb}(C) \,
\ee
with some 1-form $U_{\ga\gb}(C)= U_{\gb\ga}(C)$,
that transforms the equation (\ref{lin E}) to the vacuum one
$D_0 W_{1\,\ga\ga}'=0$. Obviously, it is possible if
\be
\label{D0 U}
   D_0 (\psi U_{\ga\gb}(C)) = {\cal T}_{\ga\gb}(C) \,,
\ee
i.e. if  ${\cal T}_{\ga\gb}(C)$ is $D_0$-exact.
Since ${\cal T}_{\ga\gb}(C)$ is $D_0$-closed
due to (\ref{lin E}) and the property $D_0^2=0$, which holds as a consequence
of the vacuum equation $R_{0\;\ga\gb}=0$, we arrive at the standard
cohomological problem. The fact that the appropriate field
redefinition exists implies that ${\cal T}_{\ga\gb}(C)$
belongs to the trivial cohomology class.

To study a question whether ${\cal T}_{\ga\gb}(C)$ is
cohomologically trivial in the class
of powers series involving arbitrary high derivatives
of the matter field $C$
we take into account that the multispinors $C_{\ga(2n)}$, $n\ge 0$,
are uniquely related to all on-mass-shell nontrivial space-time
derivatives of $C$ (for example,
$C_{\ga(4)}=-4h^\mu_{\ga\ga}h^\nu_{\ga\ga}D_{0\mu}^L
D_{0\nu}^L C$). Since ${\cal T}_{\ga\gb}(C)$ is bilinear in $C$
we can look for $U_{\ga\gb}(C)$ in the form
\begin{eqnarray}
\label{U}
   U^{\ga\ga}(C) & = & \gs\,h^{\ga\ga} C^2+
          \tau\,h^{\ga}{}_\gga C^{\gga\ga}C +
          \sum_{n=0}^\infty \left[\ga_n\, h_{\gga\gga}
          C^{\gga\gga\ga\ga\gl(2n)} C_{\gl(2n)}
              \right.  \nonumber \\
     & &  {}+\gb_n\, h_{\gga\gga}
          C^{\gga\gga\ga\gl(2n-1)} C^\ga{}_{\gl(2n-1)}
          +\gga_n\, h_{\gga\gga} C^{\gga\gga\gl(2n)}
               C^{\ga\ga}{}_{\gl(2n)}  \nonumber \\
     & &  \left.{}+\gd_n\, h_{\gga\gga}
           C^{\gga\ga\ga\gl(2n-1)} C^\gga{}_{\gl(2n-1)}
           + \gl_n\,h_{\gga\gga} C^{\gga\ga\gl(2n)}
               C^{\gga\ga}{}_{\gl(2n)} \right]   \,,
\end{eqnarray}
where $\gb_0=\gd_0=0$.
To prove that this is the most general Lorentz covariant form of
a space-time 1-form  bilinear in $C_{\ga(2n)}$
one has to use a simple property of the two-component spinors
that the total antisymmetrization with respect to any three spinor
indices gives zero. This is expressed by the identity
$$
   a_\ga(b_\gb c^\gb)
   +b_\ga(c_\gb a^\gb )+c_\ga(a_\gb b^\gb )=0  \,,
$$
which is true for any three two-component spinors
$a_\ga\,,\,b_\ga$ and $c_\ga$.
This identity allows one to express the terms of the type
$h^{\ga\ga}C^{\gl(2n)}C_{\gl(2n)}$,
$h^\ga{}_\gga C^{\gga\gl(2n-1)}C^\ga{}_{\gl(2n-1)}$,
etc. via some combinations of the terms contained in (\ref{U}).

Note that a solution of (\ref{D0 U}) is not unique due to
the ambiguity in exact shifts
$ U_{\ga\gb} \rightarrow U_{\ga\gb}+D_0(\psi V_{\ga\gb}) $,
where $V_{\ga\gb}$=$V_{\gb\ga}$
is an arbitrary 0-form bilinear in $C_{\ga(2n)}$.
The most general form of $V_{\ga\gb}$ is
\be
\label{V}
   V_{\ga\gb}=\sum_{n=0}^\infty \rho_n\,\,C^{\ga\gb\gl(2n)}
      C_{\gl(2n)}
\ee
with arbitrary coefficients $\rho_n$. An interesting fact is that
there exists an additional one-parametric ambiguity of the solutions
of this system (we hope to come back to this issue in a future publication).

Substituting (\ref{U}) into (\ref{D0 U}) and making use of
(\ref{chainb}) and (\ref{cov D0}), we arrive at
some system of algebraic equations for the coefficients
$\ga_n,\gb_n,\gga_n,\gd_n,\gl_n,\gs,\tau$.
One can fix the ambiguities in exact shifts by setting
$$
   \gl_n = 0\,,\quad n\ge 0\,,\qquad \tau=0\,,
$$
and the afore mentioned one parametric ambiguity by setting $\ga_0=0$.
Then, the comparison with the explicit form of
${\cal T}_{\ga\gb}(C)$ leads to the following set of relations
$$
   \ga_{n+1}={1\over (2n+7)(2n^2+9n+9)\gl^2}\;\left[{(2n+5)\over 2}
     \;\ga_n-{(4n^2+14n+11)\over (n+1)(2n+5)}\; \gga_n \right] \,,
     \quad n\ge 1 \,,
$$
$$
   \gb_{n+1}={1\over 2i\gl (2n^2+9n+9)} \left[
     (4n+7)\gga_n-\ga_n \right]\,, \quad n\ge 1 \,,
$$
$$
   \gga_{n+1}={1\over 2(2n+5)(n+1)\gl^2}\;\gga_n\,, \quad n\ge 1 \,,
$$
$$
   \gd_n=-i\gl\left[(2n+5)\ga_n+(2n+1)\gga_n\right]\,,
       \quad n\ge 1 \,,
$$
$$
    \ga_1=-{47\over 3360 \gl^3} \,,\quad \gb_1 = {1\over 12i\gl^2} \,,\quad
    \gga_1={3\over 160 \gl^3} \,, \quad  \gd_1={i\over 24\gl^2} \,, \quad
    \gs=-{3\over 8}\gl \,,\quad  \gga_0={3\over 8}\gl \,,
$$
which admits a unique solution. It is therefore shown that
${\cal T}_{\ga\gb}(C)$
is $D_0$-exact in the class of power series in higher derivatives.
{}From the above equations it follows that the coefficients
in front of higher derivatives acquire negative powers in $\gl$,
in agreement with the qualitative analysis in the beginning of this appendix.
In fact, the constructed solution gives a concrete realization of the
formula (\ref{nonl h}) in terms of the combinations of higher derivatives
$C_{\ga (n)}$. The similar phenomenon is expected to happen for
higher spins. The formulae obtained in this appendix have a structure
expected from the application of the integrating flow described
in sect.~\ref{Integr} in the second order in fields.


\baselineskip=9pt
\begin{thebibliography}{77}
\bibitem{hist1} C.~Fronsdal, {\it Phys. Rev.\/} {\bf D18} (1978) 3624;
         {\bf D20} (1979) 848.
\bibitem{hist2}
    J.~Fang and C.~Fronsdal, {\it Phys. Rev.\/} {\bf D18} (1978) 3630;
      {\bf D22} (1980) 1361.
\bibitem{hist3}
    P.~van~Nieuwenhuizen, {\it Phys. Rep.\/} {\bf 68} (1981) 189.

\bibitem{AD1} C.~Aragone and S.~Deser,
        {\it Phys. Lett.} {\bf B86} (1979) 161.
\bibitem{AD2}
    F.~A.~Berends, J.~W.~van ~Holten, P.~van~Niewenhuizen, and B.~de~Wit,
        {\it J. Phys.}  {\bf A13} (1980) 1643.
\bibitem{AD3}
    B.~de~Wit and D.~Z.~Freedman, {\it Phys. Rev.} {\bf D21} (1980) 358.

\bibitem{BBB1} A.~K.~Bengtsson, I.~Bengtsson, and L.~Brink, {\it Nucl. Phys.\/}
      {\bf B227} (1983) 31, 41.
\bibitem{BBB2}
   F.~A.~Berends, G.~J.~Burgers, and H.~van~Dam, {\it Z. Phys.\/}
      {\bf C24} (1984) 247;
      {\it Nucl.~Phys.\/} {\bf B260} (1985) 295; {\bf B271} (1986) 429.
\bibitem{BBB3}
    A.~K.~H.~Bengtsson and I.~Bengtsson, {\it Class. Quant. Grav.\/}
      {\bf 3} (1986) 927.
\bibitem{BBB4}
    A.~K.~H.~Bengtsson, {\it Class. Quant. Grav.\/} {\bf 5} (1988) 437.
\bibitem{BBB5}
    R.~Metsaev, {\it Mod. Phys. Lett.\/} {\bf A6} (1991) 359.
\bibitem{nogo1} S.~Coleman and J.~Mandula, {\it Phys. Rev.}
      {\bf 159} (1967) 1251.
\bibitem{nogo2}  R.~Haag, J.~Lopuszanski, and M.~Sohnius,
    {\it Nucl. Phys.\/} {\bf B88} (1975) 257.
\bibitem{FV1} E.~S.~Fradkin and M.~A.~Vasiliev, {\it Phys. Lett.\/}
    {\bf B189} (1987) 89; {\it Nucl.~Phys.\/} {\bf B291} (1987) 141.
\bibitem{more} M.~A.~Vasiliev, {\it Phys.~Lett.} {\bf B285} (1992) 225
     (and references therein).
\bibitem{rev} M.~A.~Vasiliev, {\it Int.~J.~Mod.~Phys.} {\bf D5} (1996) 763.
\bibitem{AdS1} J.~Maldacena,
    ``The Large $N$ Limit of Superconformal Field Theories
      and Supergravity", hep-th/9711200.
\bibitem{FF} S. Ferrara and C. Fronsdal,
``Conformal Maxwell Theory as a Singleton Field Theory on $ADS_5$,
IIB Branes and Duality",
hep-th/9712239.
\bibitem{AdS2} M.~Gunaydin and D.~Minic, ``Singletons, Doubletons and
      $M$-theory", hep-th/9802047.
\bibitem{AdS3} S.~S.~Gubser, I.~R.~Klebanov, and A.~M.~Polyakov,
     ``Gauge Theory Correlators from Non-Critical String Theory",
      hep-th/9802109.
\bibitem{AdS4} E.~Witten, ``Anti De Sitter Space and Holography",
      hep-th/9802150.
\bibitem{singl1} P.~A.~M.~Dirac, {\it J. Math. Phys.} {\bf 4} (1963) 901.
\bibitem{singl2} M.~Flato and C.~Fronsdal,
   {\it Lett.~Mat.~Phys.}~{\bf 2} (1978) 421;
   {\it Phys.~Lett.}~{\bf B97} (1980) 236.
\bibitem{vafa} C.~Vafa, ``Puzzles at Large $N$", hep-th/9804172.
\bibitem{sez}  E.~Sezgin and P.~Sundell, ``Higher Spin $N=8$ Supergravity",
  hep-th/9805125.
\bibitem{W} E.~S.~Fradkin and M.~A.~Vasiliev, {\it Dokl. Acad. Nauk.}
   {\bf 29} (1986) 1100; {\it Ann. of Phys.} {\bf 177} (1987) 63.
\bibitem{OP1} M.~A.~Vasiliev, {\it Fortschr. Phys.} {\bf 36} (1988) 33.
\bibitem{KV1} S.~E.~Konstein and M.~A.~Vasiliev, {\it Nucl. Phys.\/}
  {\bf B331} (1990) 475.
\bibitem{OP2} M.~A.~Vasiliev, {\it JETP Lett.} {\bf 50} (1989) No.8, 374;
     {\it Int.~J.~Mod.~Phys.} {\bf A6} (1991) 1115.
\bibitem{MOY1} J.~Moyal, {\it Proc. Camb. Phil. Soc.} {\bf 45} (1949) 99.
\bibitem{MOY2} F.~Bayen, M.~Flato, C.~Fronsdal, A.~Lichnerowicz
    and D.~Sternheimer, {\it Ann. Phys.} (N.Y.) {\bf 110} (1978) 61, 111.
\bibitem{MM1} A.~Connes, M.~R.~Douglas and A.~Schwarz, ``Noncommutative
  Geometry and Matrix Theory: Compactification on Tori", hep-th/9711162.
\bibitem{MM2}  M.~Douglas and C.~Hull, ``D-Branes and Noncommutative Torus",
  hep-th/9711165.
\bibitem{MM3}  M.~Berkooz,
  ``Nonlocal Field Theories and the Noncommutative Torus", hep-th/9802069.
\bibitem{MM4}  N.~Nekrasov and A.~Schwarz,
  ``Instantons on noncommutative ${\bf R^4}$
  and (2,0) superconformal six dimensional theory", hep-th/9802068.
\bibitem{Castro} C.~Castro, ``$W$ Geometry {}From Fedosov's Deformation
      Quantization", \\hep-th/9802023.
\bibitem{DL} C.~Devchand and O.~Lechtenfeld, ``Extended Self-Dual Yang-Mills
     from the $N=2$ String",  {\it Nucl.~Phys.}~{\bf B516} (1998) 255,
     hep-th/9712043.
\bibitem{AchTow} A.~Achucarro and P.~K.~Townsend, {\it Phys.~Lett.}~{\bf B180}
     (1986) 89.
\bibitem{Witt} E.~Witten,~{\it Nucl.~Phys.}~{\bf B311} (1989) 46.
\bibitem{Blen} M.~P.~Blencowe, {\it Class. Quantum Grav.}~{\bf 6}
  (1989) 443.
\bibitem{Eq} M.~A.~Vasiliev, {\it Mod.~Phys.~Lett.} {\bf A7} (1992) 3689.
\bibitem{Nic} H.~Nicolai, {\it Nucl.~Phys.}~{\bf B176} (1980) 419.
\bibitem{Unf} M.~A.~Vasiliev, {\it Class.~Quant.~Grav.} {\bf 11} (1994) 649.
\bibitem{BPV} A.~V.~Barabanschikov, S.~F.~Prokushkin, and
   M.~A.~Vasiliev, {\it Theor.~Math.~Phys.} {\bf 110} (1997) 295,
   hep-th/9609034.
\bibitem{geom1} R.~Utiyama, {\it Phys.~Rev.} {\bf D101} (1956) 1597.
\bibitem{geom2} T.~W.~B.~Kibble, {\it J.~Math.~Phys.} {\bf 2} (1961) 212.
\bibitem{geom3} A.~H.~Chamseddine and P.~C.~West, {\it Nucl.~Phys.}
    {\bf B129} (1977) 39.
\bibitem{geom4} S.~W.~MacDowell and F.~Mansouri, {\it Phys.~Rev.~Lett.}
    {\bf 38} (1977) 739.
\bibitem{geom5} K.~Stelle and P.~West, {\it Phys.~Rev.} {\bf D21} (1980) 1466.
\bibitem{Hawk}  S.~W.~Hawking and D.~Page, ``Thermodynamics of Black Holes In
    Anti-de Sitter Space", {\it Commun. Math. Phys.} {\bf 87} (1983) 577.
\bibitem{Wig} E.~P.~Wigner, {\it Phys.~Rev.} {\bf 77} (1950) 711.
\bibitem{bershub}  F.~A.~Berezin and M.~A.~Shubin,
      {\it ``Schr\"odinger Equation"}, Moscow Univ. Press, Moscow, 1983.
\bibitem{Prop} M.~A.~Vasiliev, {\it Class.~Quantum~Grav.} {\bf 8} (1991) 1387.
\bibitem{Tbilisi} M.~A.~Vasiliev, Proceedings of the Second International
      Workshop on Selected Topics of Theoretical and Modern Mathematical
      Physics, Tbilisi, Georgia, September 22-28, Eds: M.~Eliashvili,
      G.~Jorjadze, E.~Ragoucy and P.~Sorba, ENSLAPP-A-642/97.
\bibitem{buck1} M.~A.~Vasiliev, {\it Nucl.~Phys.~B~(Proc.~Suppl.)}
      {\bf 56B} (1997) 241.
\bibitem{vol} M.~A.~Vasiliev,
     ``Deformed Oscillator Algebras and Higher Spin Gauge Interactions
      of Matter Fields in 2+1 Dimensions", hep-th/9712246.
\bibitem{kir} D.~A.~Kirzhnitz, {\it JETP} {\bf 49} (1965) 1544;
    in {\it Sbornik ``Problems of Theoretical Physics" in memory
    of I.~E.~Tamm}, Moscow, ``Nauka", 1972.
\bibitem{Ann} M.~A.~Vasiliev, {\it Ann. Phys.} (N.Y.) {\bf 190} (1989) 59.
\bibitem{GSW} M.~Green, J.~Schwarz, E.~Witten,
    {\it ``Superstring Theory"}, vols. 1 and 2, Cambridge Univ. Press,
     New York, 1987.

\bibitem{FV2} E.~S.~Fradkin and M.~A.~Vasiliev,
    {\it Int.~J.~Mod.~Phys.}~{\bf A3} (1988) 2983.
\bibitem{BWV} E.~Bergshoeff, B.~de~Wit, and M.~A.~Vasiliev,
              {\it Nucl.~Phys.} {\bf B366} (1991) 315.
\bibitem{dwtol} B.~de~Wit, A.~K.~Tollst\'en, and H.~Nicolai,
              {\it Nucl.~Phys.} {\bf B392} (1993) 3, hep-th/9208074.

\bibitem{SezNic} H.~Nicolai, E.~Sezgin, and Y.~Tanii,
      {\it Nucl.~Phys.} {\bf B305} [FS23] (1988) 483.

\bibitem{sprav} E.~Janke, F.~Emde, F.~L\"osh,
      {\it Tafeln H\"oherer Funktionen}, B.~G.~Teubner
      Verlagsgesellsehaft, Stuttgart, 1960.

\end{thebibliography}
\end{document}